\documentclass[aps,preprint,12pt,eqsecnum,nofootinbib,amsmath,amssymb]{revtex4}

\newcommand\mydates{25 November 2007}

\usepackage{graphicx} 
\usepackage{dcolumn}  
\usepackage{bm}       
\usepackage{latexsym} 
\usepackage{fancyhdr} 

\newcommand{\titleskip}{\baselineskip 18pt plus 1pt minus 1pt}
\newcommand{\affiliationskip}{\baselineskip 15pt plus 1pt minus 1pt}
\newcommand{\abstractskip}{\baselineskip 13pt plus 1pt minus 1pt}
\newcommand{\tableofcontentsskip}{\baselineskip 14pt plus 1pt minus 1pt}
\newcommand{\bodyskip}{\baselineskip 18pt plus 1pt minus 1pt}
\newcommand{\footnoteskip}{\baselineskip 12pt plus 1pt minus 1pt}
\newcommand{\captionskip}{\footnotesize \baselineskip 12pt plus 1pt minus 1pt}

\newcommand{\mmbf}[1]{\mbox{\boldmath${#1}$}}        

\newcommand{\smPerp}{{\scriptscriptstyle \perp }}
\newcommand{\smGT}{{\scriptscriptstyle >}}
\newcommand{\smLT}{{\scriptscriptstyle <}}

\newcommand{\smBPS}{{\rm\scriptscriptstyle BPS}}

\newcommand{\smD}{{\rm\scriptscriptstyle D}}
\newcommand{\smI}{{\rm\scriptscriptstyle I}}

\newcommand{\smB}{{\rm\scriptscriptstyle B}}

\newcommand{\smCoul}{{\rm\scriptscriptstyle coul}}
\newcommand{\smTimes}{{\scriptstyle \times}}

\begin{document}

\preprint{\hfill LA-UR-07-2173}

\title{\titleskip
  BPS Explained II: Calculating the Equilibration \\
  Rate in the Extreme Quantum Limit
}

\author{Robert~L. Singleton Jr.}

\vskip0.2cm 
\affiliation{\affiliationskip
     Los Alamos National Laboratory\\
     Los Alamos, New Mexico 87545, USA
}
\date{\mydates}

\begin{abstract}
\abstractskip
\vskip0.3cm 

\noindent

This is the second in a series of two lectures on the technique of
dimensional continuation, a new method for analytically calculating
certain energy transport quantities in a weakly to moderately coupled
plasma.  Recently, this method was employed by Brown, Preston, and
Singleton (BPS) to calculate the electron-ion temperature
equilibration rate and the charged particle stopping power to leading
and next-to-leading order in the plasma coupling. The basic idea is
very simple. Concentrating upon the equilibration rate, the
calculation consists of the following two steps: (i)~perturbatively
expand the rate in the form \hbox{$d { \cal E }/ dt = - A\, g^2 \ln g
+ B g^2 + {\cal O}(g^3)$}, with the dimensionless expansion parameter
being defined by $g=e^2 \kappa_e/4\pi T_e$; (ii) analytically
calculate the coefficients $A$ and $B$ using the method of dimensional
continuation.  The factor of $4\pi$ should be omitted from $g$ in
nonrationalized electrostatic units.  In the first lecture, I
presented a basic overview of the requisite theoretical machinery of
dimensional continuation imported from particle physics, but in a
self-contained manner that assumed no familiarity with quantum field
theory. In this lecture, I develop the framework further, and then
explicitly calculate the electron-ion temperature equilibration rate
in the high temperature limit. In this extreme quantum limit, the
calculation of the coefficients $A$ and $B$ simplifies considerably,
allowing us to concentrate on the physics of the method rather than
the added complexity of the more general BPS calculation. This method
captures {\em all} short and long distance physics to second order in
$g$, while three-body and higher correlations are contained in the
cubic and higher order terms denoted by ${\cal O}(g^3)$. In a weakly
to moderately coupled plasma, where $g$ is small, the error term
${\cal O}(g^3)$ in this calculation is also small compared to the $A$-
and $B$-terms, in which case the BPS methodology is quite
accurate. Should higher order contributions be required, they can be
calculated systematically, thereby improving the accuracy of the
result in a controlled manner. To get a feel for the numbers, one
finds $g \sim 0.04$ at the center of the sun, where the plasma
conditions are $n \sim 5 \times 10^{25}\,{\rm cm}^{-3}$ and $T \sim
1\,{\rm keV}$. The coupling constant $g$ can be scaled to other plasma
regimes through the proportionality relation $g\propto n^{1/2}\,
T^{-3/2}$.  Of course the application of interest determines the
relevant plasma regime, which may or may not lie within the domain of
applicability of the BPS calculation. For example, the technique
breaks down for warm dense matter where is $g$ not very small;
however, this analytic perturbative technique is applicable for
ignition in inertial confinement fusion and for other processes in hot
a weakly coupled plasma.

\end{abstract}

\maketitle

\pagebreak
\tableofcontentsskip
\tableofcontents
\thispagestyle{empty}

\pagebreak
\bodyskip
\setcounter{page}{1}

%
\section{Introduction and Review}

This is the second lecture on dimensional continuation, a new
technique~\cite{lfirst} recently used to calculate the charged
particle stopping power and the temperature equilibration rate in a
weakly to moderately coupled plasma~\cite{bps}. In
Lecture~I~\cite{bps1} of this series, I discussed the basic
theoretical machinery of dimensional continuation, and I performed a
model calculation of the equilibration rate. Reference~\cite{degen} also
contains a summary of the method in a very readable form.  In this
lecture, I will present the complete calculation of the electron-ion
temperature equilibration rate in the {\em extreme} quantum limit,
valid to leading and next-to-leading order in the number density (a
more general calculation is performed in Section~12 of Ref.~\cite{bps}
to {\em all} orders in quantum mechanics, thereby providing an exact
interpolation between the extreme classical and quantum limits). This
calculation is near exact for a weakly coupled plasma, and it is quite
accurate for a moderately coupled plasma.  Before proceeding directly
to the calculation, however, it might be useful to quickly review some
of the more salient features of dimensional continuation discussed in
Lecture~I.

Under most circumstances, a plasma is not produced in thermal
equilibrium; for example, when a laser ionizes a substance, it
preferentially heats the electrons over the ions. However, since the
electrons are so light, they rapidly come into thermal equilibrium
among themselves with temperature $T_e$; some time later, the ions too
will equilibrate among themselves to a common temperature
$T_\smI$. Finally, the electrons and ions will begin to equilibrate,
and it is this process upon which we shall focus. Let $d{\cal E}_{e
\smI}/dt$ denote the rate per unit volume at which the electron system
at temperature $T_e$ exchanges energy with the ion system at
temperature $T_\smI$ through Coulomb interactions (throughout these
notes, I will always measure temperature in energy units).  The
electron-ion equilibration rate is proportional to the temperature
difference between the electrons and ions, and can be expressed by
\begin{eqnarray}
  \frac{d{\cal E}_{e\smI}}{dt}
  =   -\, {\cal C}_{e\smI}\left(T_e-T_\smI\right) \ .
\label{dedteI}
\end{eqnarray}
To restate the goal of this lecture more precisely, we shall calculate
${\cal C}_{e\smI}$ in the high temperature limit [where two-body
scattering is accurately given by the Born approximation], and we will
do so {\em exactly} to leading and next-to-leading order in the plasma
coupling parameter $g$ [defined in Lecture~I, or in Eq.~(\ref{gedef})
of this lecture].  Under these conditions, the result takes a
particularly simple form~\cite{bps}:
\begin{eqnarray}
  {\cal C}_{e\smI}
  = 
  \frac{\omega_\smI^2}{2\pi}\, \kappa_e^2\,
  \sqrt{\frac{m_e}{2\pi\, T_e}}\, \ln\Lambda_\smBPS \ ,
  ~~~\text{with}~~~
  \ln\Lambda_\smBPS
  =
  \frac{1}{2}\left[\ln\!\left\{\frac{8 T_e^2}{\hbar^2 \omega_e^2}
  \right\} - \gamma - 1 \right] \ ,
\label{bpsrate}
\end{eqnarray}
\vskip0.5cm \noindent
where $\gamma=0.57721 \cdots$ is the Euler constant, $\kappa_e$ and
$\omega_e$ are the electron Debye wave number and plasma frequency,
and \hbox{$\omega_\smI^2 =\sum_i \omega_i^2$} is sum of the squares of
the ion plasma frequencies.\footnote{  \footnoteskip
  Equation~(\ref{bpsrate}) corresponds to
  Eqs.~(3.61)~and~(12.12) of Ref.~\cite{bps}, where I have taken this
  opportunity to correct a small transcription error: when passing from
  Eq.~(12.43) to Eq.~(12.44) in Ref.~\cite{bps}, a factor of 1/2 was
  dropped. Restoring this factor of 1/2 changes the additive constant
  outside the logarithm from the $-\gamma-2$ that appears in Eq.~(12.12)
  of Ref.~\cite{bps} to the constant $-\gamma-1$ in~(\ref{bpsrate}).
}

In the form displayed by equation (\ref{bpsrate}), the rate
coefficient ${\cal C}_{e \smI}$ and the Coulomb logarithm $\ln
\Lambda_\smBPS$ do not explicitly depend upon one's choice of
electrostatic units, and one may calculate the Debye wave numbers and
the plasma frequencies in any desired system.  For dimensional
continuation, however, it is more convenient to use {\em rationalized}
electrostatic units, and I shall employ this choice from here out.  An
arbitrary plasma component will be labeled by an index $b$, and is
characterized by mass $m_b$, charge $e_b=Z_b\, e$, number density
$n_b$, and temperature $T_b$. The index $b$ can span the electron and
ion plasma components, that is to say, $b=e,i$ with $i$ being an
arbitrary ion species.  Working in three dimensions for now, the
Coulomb potential between two charges $e_a$ and $e_b$ separated by a
distance $r$ is \hbox{$V=e_a e_b/4\pi \,r$}, and in rationalized
units, the Debye wave number and the plasma frequency of species $b$
take the form\footnote{\footnoteskip
  In nonrationalized units, the right-hand-sides of (\ref{kbdef}) and
  (\ref{wbdef}) should contain an additional factor of $4\pi$.
}
\begin{eqnarray}
  \kappa_b^2 &=& \frac{e_b^2\, n_b}{T_b}
\label{kbdef}
\\[5pt]
  \omega_b^2 &=& \frac{e_b^2\, n_b}{m_b} \ .
\label{wbdef}
\end{eqnarray}
The square of the total Debye wave number is $\kappa_\smD^2
=\sum_b \kappa_b^2$, and the total Debye wave length is $\lambda_\smD
= \kappa_\smD^{-1}$.

\section{Calculating the Rate in Perturbation Theory}

Reference~\cite{bps}, hereafter referred to as BPS, used a double
pronged strategy to calculate the rate coefficient (\ref{bpsrate}).
First, a well chosen~\cite{by} dimensionless parameter $g$ was
constructed from the relevant dimensionfull plasma quantities, thereby
providing a parameter in which to perform a {\em controlled}
perturbative expansion to leading and next-to-leading order in
$g$. The systematic error of the calculation was estimated by the
cubic order term in the expansion, which is quite small for a weakly
to moderately coupled plasma.  While perturbative calculations are not
very common in plasma physics, primarily because of the complexity of
the systems of interest and the computational focus within the field,
the validity of perturbation theory should nonetheless be clear for a
``simple'' system such as a weakly coupled and fully ionized plasma.
The second part of the BPS argument deployed a powerful technique from
quantum field theory allowing one to {\em analytically} calculate
the coefficients in the $g$-expansion. 

\subsection{Perturbative Expansions in Weakly Coupled Plasmas}

Let us first concentrate on the perturbative expansion. As
demonstrated in Ref.~\cite{by}, and discussed at length in
Lecture~I~\cite{bps1}, for the case at hand the dimensionless plasma
coupling parameter is defined by\footnote{
\footnoteskip
  In nonrationalized units we would write $g = e^2 \kappa_e/T_e$, with
  $\kappa_e^2 = 4\pi\, e^2 n_e/T_e$. 
}
\begin{eqnarray}
  g &=& \frac{e^2\, \kappa_e}{4\pi T_e} \ .
\label{gedef}
\end{eqnarray}
Note that $g$ is the ratio of the Coulomb potential energy of two
point-charges, separated by the screening length $\kappa_e^{-1}$, to
the thermal energy of the plasma. Therefore, $g$ can be used to
measure the strength of the plasma. To get a feel for the size of $g$
in a hot but not too dense plasma, one finds $g = 0.042$ for a
hydrogen plasma under the \hbox{solar-like} conditions $n_e = 5.0
\times 10^{25}\,{\rm cm}^{-3}$ and $T_e = 1.3\,{\rm keV}$. One can
scale to other density and temperature regimes by noting that
$g\propto n_e^{1/2}\, T_e^{-3/2}$.  It was shown in Ref.~\cite{by}
that plasma quantities always expand in {\em integer} powers of the
coupling $g$, and therefore $g$ is the appropriate parameter in which
to perform a controlled perturbative analysis for weakly coupled
plasmas.\footnote{\footnoteskip 
  The usual plasma parameter $\Gamma$ is related to the expansion
  parameter by $g \propto \Gamma^{3/2}$ (with proportionality constant
  of order unity). Small values of $\Gamma$ therefore imply small
  values of $g$, and one may characterize the {\em strength} of the
  plasma by either $g$ or $\Gamma$. The proportionality relation above
  follows from the fact that $g\propto n_e^{1/2}$ and $\Gamma \propto
  n_e^{1/3}$ (the parameter $g$ is defined in terms of a Debye
  screening length $\kappa_e^{-1}$, while $\Gamma$ is defined in terms
  of the inter-particle spacing $n_e^{-1/3}$).  Furthermore, since $g
  \propto n_e^{1/2}$, we may loosely think of the $g$-expansion as an
  expansion in the electron number density, as I have done in the
  first paragraph of this introduction. More precisely, of course, we
  are expanding in the dimensionless quantity $g \propto e^3\,
  n_e^{1/2} \, T_e^{ -3/2}$.  See Ref.~\cite{by} for more details,
  particularly Section~1.1 entitled {\em Relevant Scales and
  Dimensionless Parameters}.  }
The \hbox{$g$-expansion} allows for possible non-analytic terms, such
as $\ln g$, and in particular, the electron-ion equilibration rate can
be written
\begin{eqnarray}
  \frac{d{\cal E}_{e\smI}}{dt} 
  = 
  -\underbrace{A\, g^2\ln g}_\text{LO}
  \,+\, 
  \underbrace{~B g^2~}_\text{NLO} \,+\,\,  {\cal O}(g^3) \ ,
\label{dedtNLO}
\end{eqnarray}
where I have indicated the leading order (LO) and the next-to-leading
order (NLO) terms in the expansion. The minus sign on the leading
order term of (\ref{dedtNLO}) is a matter of convention, and for small
values of $g$ it renders
the coefficient $A$ positive when the energy exchange is positive.
Provided we can calculate the coefficients $A$ and $B$, then
(\ref{dedtNLO}) will be quite accurate in a weakly to moderately
coupled plasma in which $g$ is small. 
Of course this perturbative approach breaks
down for strongly coupled plasmas, those for which the value of $g$ is
of order one or greater, since every term in the expansion becomes
equally important in such cases. However, unlike a model or an
uncontrolled calculation, the BPS calculation informs us of its domain
of validity, and it provides an estimate of its own error through
the size of $g$.

\subsection{Calculating the Coefficients of the Expansion}

We have now reduced the problem to finding the coefficients 
$A$ and $B$ of the rate (\ref{dedtNLO}).  The coefficient $A$ was
first obtained long ago by Spitzer~\cite{sbook}~(and it can be
estimated by dimensional analysis alone). The coefficient $B$,
however, was calculated only recently in Ref.~\cite{bps}, which
employed a powerful technique from quantum field theory called
dimensional regularization, or dimensional continuation as I will call
it here. Since this technique is quite subtle and has proven to be
somewhat controversial, I should emphasize that the method by which
one {\em chooses} to calculate these coefficients is immaterial,
except to the extent that it must contain enough physics to extract
the next-to-leading order coefficient $B$.  Techniques other than
dimensional continuation could well furnish one with the correct
expressions for $A$ and $B$, and perhaps in a simpler manner. However,
the only relevant point here is that~\hbox{\em by hook or by crook} we
must analytically calculate these coefficients, and dimensional
continuation is one method of doing this.\,\footnote{\footnoteskip
  I have recently been informed\,\cite{jcom} that Ref.~\cite{jm},
  which is designed to apply to both strongly and weakly coupled
  plasmas, reproduces the BPS result (\ref{bpsrate}) in the limit of
  weak coupling. As far as I am aware, Refs.~\cite{bps} and \cite{jm}
  are the only works currently in the literature with a formalism
  strong enough to extract such next-to-leading order physics from
  first principles.
} 

Before turning to the calculation of the coefficients,
allow me to make a comment on the relation between the next-to-leading
order $B$-term and the Coulomb logarithm.  Writing the leading order
coefficient as $K=A g^2$, and defining the dimensionless coefficient
$C=\exp\{-B/A\}$, we can express the rate (\ref{dedtNLO}) in the form
\begin{eqnarray}
  \frac{d{\cal E}_{e\smI}}{dt} 
  &=& 
  K\ln\Lambda_\smCoul \,+\, {\cal O}(g^3) \ ,
  ~~~\text{with}~~~
  \ln\Lambda_\smCoul = -\ln\left\{ C g\right\} \ .
\label{lngsqu}
\end{eqnarray}
Since the Coulomb logarithm means different things to different
people,\footnote{\footnoteskip
  Student: What is the Coulomb logarithm? Professor: 10.  }
I would like to be quite specific in this lecture. By the words
``Coulomb logarithm'' I simply mean the term $\ln\!\Lambda_\smCoul$
defined in (\ref{lngsqu}), excluding the cubic and higher order terms.
Hence, calculating the next-to-leading order coefficient $B$ is
equivalent to determining the dimensionless coefficient $C$ {\em
inside} the Coulomb logarithm. Finding dimensionless constants is
usually a difficult problem, particularly since one cannot appeal to
dimensional analysis for an estimate. It should therefore not be
surprising that the coefficient $C$ varies over an order of magnitude
or so across the various models within the literature.

\vfill
\pagebreak
\section{\label{sec:framework} Calculating in Arbitrary Dimensions}

Before proceeding directly to the calculation in
Sec.~\ref{sec:calmain}, let us further develop the basic physics and
mathematical machinery necessary to perform calculations in an
arbitrary number of dimensions. The motivation for this section is, of
course, a thorough exposition of the BPS methodology for calculating
Coulomb energy-loss processes in a plasma. However, the material in
this section is well known and applicable to a wide variety of other
calculations, such as particle decay rates in high energy physics and
analytic work in statistical mechanics.  For the sake of completeness,
however, and to establish some results that will be useful in
Sec.~\ref{sec:calmain}, I will present a cursory but self-contained
treatment here. If this material is familiar, then one may proceed
directly to the calculation of the temperature equilibration rate in
Sec.~\ref{sec:calmain} (given the background material in this section,
the calculation itself is less than eight pages in length).

We shall start by developing the hyperspherical coordinate system in
$\nu$ dimensions, which is a straightforward generalization of three
dimensional spherical coordinates. To illustrate the utility of
hyperspherical coordinates, I will calculate the hyperarea and
hypervolume of several multidimensional objects by exploiting their
spherical and cylindrical symmetries. These results will be used quite
extensively in the next section. As a physical application, I then
develop the multidimensional analog of the scattering cross section,
which will allow us to consistently include short-distance quantum
scattering effects in the $g$-expansion (quantum effects manifest
themselves through the $\eta$-dependence of the coefficients in this
expansion). Since we are interested in Coulomb energy exchange, we
next examine electrostatics in arbitrary dimensions. From the
multidimensional form of Gauss' Law, we shall derive the
$\nu$-dimensional Coulomb potential $V_\nu({\bf x})$, and we will see
that it depends only upon $r=\vert {\bf x}\vert$ in such a way as to
emphasizes short distance physics when $\nu>3$ and long distance
physics when $\nu<3$.  In $\nu=3$, the short and long distance physics
compete with equal strength, giving an infrared and an ultraviolet
divergence, and this is what renders the temperature equilibration
problem so difficult. To employ the extreme quantum limit, in which
the Born approximation for the two-body scattering dominates, we must
calculate the Fourier transform of the Coulomb potential in $\nu$
dimensions. Interestingly, we shall find that the Fourier transform of
$V_\nu(r)$ is given by the quite simple expression $\tilde V_\nu({\bf
k}) = 1/k^2$, the form of which does not depend upon the dimension of
space, but only upon the length of the wavenumber $k=\vert {\bf k}
\vert$.  The fact that $\tilde V_\nu(k)$ is so simple greatly
facilitates calculations in the extreme quantum limit. With potential
in hand, we shall then construct kinetic equations in $\nu$
dimensions. These equations are explicitly finite in all but $\nu=3$
dimensions, and I will explain the manner by which the BBGKY hierarchy
reduces to the Boltzmann equation and the Lenard-Balescu equation (in
$\nu>3$ and $\nu<3$ respectively).

\subsection{\label{sec:one} Kinematics and Hyperspherical Coordinates}

\subsubsection{\label{sec:hypcoord} Hyperspherical Coordinates}

Kinematic quantities such as the $\nu$-dimensional momentum or
position vectors are elements of the same $\nu$-dimensional Euclidean
space $\mathbb{R}^\nu$. For definiteness, I will specialize to the
case of position ${\bf r}$, with the understanding that this vector
could also refer to momentum or wavenumber.  We can decompose any
vector ${\bf r} \in \mathbb{R}^\nu$ in terms of a rectilinear
orthonormal basis $\hat{\bm e}_\ell$, so that ${\bf r}=
\sum_{\ell=1}^\nu x_\ell\,\hat{\bm e}_\ell$, or in component notation
${\bf r } = (x_1,\cdots,x_\nu)$. Each component is given by $x_\ell =
\hat{\bm e}_\ell \cdot {\bf r}$, and a change $d{\bf r}$ in the vector
${\bf r}$ corresponds to a change $dx_\ell = \hat{\bm e}_\ell \cdot
d{\bf r}$ in the rectilinear coordinate $x_\ell$. Letting ${\bf r}$
vary successively along the independent directions $\hat{\bm e}_\ell$,
we can trace out a small $\nu$-dimensional hypercube with sides of
length $dx_\ell$; therefore, the rectilinear volume element is given
by the simple form
\begin{eqnarray}
  d^\nu x 
  = 
  \prod_{\ell=1}^\nu dx_\ell
  = 
  dx_1\, dx_2 \cdots dx_\nu \ .
\label{dnu:rec}
\end{eqnarray}
In performing integrals over the kinematic variables, however,
symmetry usually dictates the use of hyperspherical coordinates rather
than rectilinear coordinates. I will therefore review the
hyperspherical coordinate system in this subsection, deriving the
measure for a \hbox{$\nu$-dimensional} volume element $d^\nu x$ in
terms of hyperspherical coordinates.  For our purposes, the primary
utility of hyperspherical coordinates is that the volume element
$d^\nu x$ can be written as a product of certain conveniently chosen
dimensionless angles, which I will collectively refer to as
$d\Omega_{\nu-1}$, and an overall dimensionfull radial factor
$r^{\nu-1}\, dr$, so that $d^\nu x = d\Omega_{\nu-1}\, r^{\nu-1}dr$. 

\begin{figure}[t]
\includegraphics[scale=0.45]{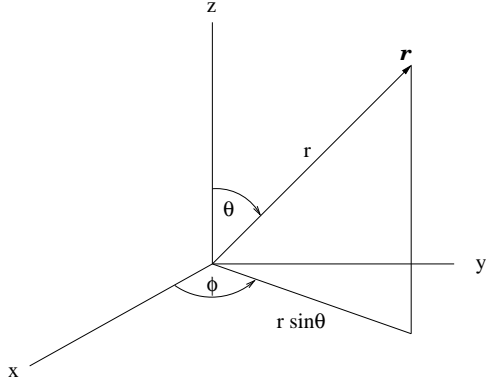}
\vskip-0.3cm 
\caption{\captionskip
  Spherical coordinates $r,\theta,\phi$ of a point ${\bf r}$ in three
  dimensional space: radial distance $r$, polar angle $\theta$, and
  azimuthal angle $\phi$. The angles range over the values $0 \le
  \theta \le \pi$ and $0 \le \phi < 2\pi$.
}
\label{fig:coord1}
\end{figure}

Starting with the usual \hbox{3-dimensional} spherical coordinates of
Fig.~\ref{fig:coord1}, let us recall why the three dimensional volume
element takes the form $d^3 x = \sin\theta\, d\theta\, d\phi\, r^2 dr$
(with \hbox{$0 \le \theta \le \pi$} and \hbox{$0 \le \phi < 2\pi$},
and of course $0 \le r < \infty$; the coordinate singularities of the
spherical system are not important here). As depicted in the figure,
the three dimensional vector ${\bf r}$ has length $r$, and subtends a
polar angle $\theta$ relative to the \hbox{$z$-axis}, while its
projection onto the \hbox{$x$-$y$ plane} subtends an azimuthal angle
$\phi$ relative to the $x$-axis.  The two angles $\theta$ and $\phi$
specify completely the direction of the unit vector $\hat{\bf r}$. As
we increase the polar angle $\theta$ by a small amount $d\theta$, the
vector ${\bf r}$ sweeps out an arc of length $dR_1 = r d\theta$;
similarly, a change $d\phi$ in the azimuthal angle will cause ${\bf
r}$ to sweep out a perpendicular arc (in the $x$-$y$ plane) of length
$dR_2 = r \sin\theta\,d\phi$. Note that the factor of $\sin\theta$ in
$dR_2$ arises from the projection of ${\bf r}$ onto the $x$-$y$
plane. We can make one more independent displacement by moving $dr$
units in the radial direction, which results in a line of length
$dR_3=dr$. For small displacements in $d\theta$, $d\phi$, and $dr$,
the vector ${\bf r}$ sweeps out a small cubic volume element with
sides of length $dR_1$, $dR_2$, and $dR_3$. The volume of this element
is therefore $d^3 x= d R_1\, d R_2\, dR_3 = r d\theta \cdot
r\sin\theta d\phi \cdot dr$.

\begin{figure}[t]
\includegraphics[scale=0.5]{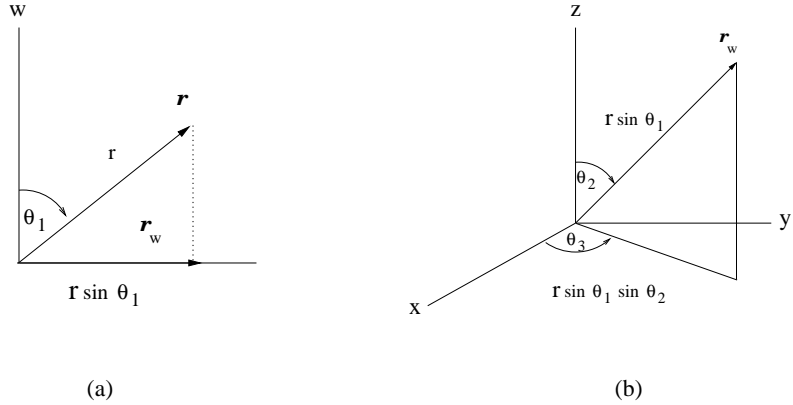}
\vskip-0.3cm 
\caption{\captionskip
  Hyperspherical coordinates $r,\theta_1,\theta_2,\theta_3$ of a point
  ${\bf r}$ in four dimensional space. As before, $r=\vert {\bf r}
  \vert$ is the radial distance. The angles are defined as follows.
  (a) First, let $\theta_1$ be the angle between ${\bf r}$ and the
  $w$-axis. Let us now project ${\bf r}$ onto the orthogonal three
  dimensional space, so that ${\bf r}=(x,y,z,w) \to {\bf r }_w =
  (x,y,z,0)$. The length of this projection is $r_w = r \sin
  \theta_1$, and the projection itself is the same as projecting ${\bf
  r}$ onto the three dimensional hyperplane $w=0$. (b) The vector
  ${\bf r}_w$ can be viewed as a three dimensional vector ${\bf
  r}_w=(x,y,z)$, which then defines the usual polar and azimuthal
  angles of Fig.~\ref{fig:coord1}, denoted here by $\theta_2$ and
  $\theta_3$ respectively.
}
\label{fig:coord3}
\end{figure}

Let us now consider the volume element $d^4x$ in four dimensional
space. Denote the coordinates of a vector ${\bf r}$ by $x, y, z, w$,
that is to say, take ${\bf r}=(x,y,z,w)$.  Since we cannot visualize
four dimensional space,\footnote{
  \footnoteskip Apart from visualization problems, we can nonetheless
  work in higher dimensions by employing analytic geometry and
  analogies with lower dimensions. For example, an ordinary two-sphere
  of radius $r$, which I will denote by $S_2(r)$, has the equation
  $x^2+y^2+z^2=r^2$ in three dimensional space; a corresponding
  ``three-sphere'' $S_3(r)$ in four dimensions can be represented by
  $x^2 + y^2 + z^2 + w^2=r^2$. In a similar manner, a two-dimensional
  plane in three-space can be expressed as $a_1\,x+ a_2\,y+ a_3\,z= c$
  for real numbers $a_\ell$ and $c$, while a three-plane in four
  dimensional space takes the form $a_1\,x+ a_2\,y+ a_3\,z + a_4\, w=
  c$. As a final example, consider a ``three-cone'' oriented along the
  $w$-axis: $x^2+y^2+z^2-w^2=0$. The ``conic sections'' are obtained
  by slicing this hypercone with a three-plane along various
  orientations. For example, if we slice the three-cone by the
  hyperplane $w=r$ orthogonal to the $w$-axis, then we find a
  two-sphere $x^2+y^2+z^2=r^2$; if we slice the three-cone by a
  hyperplane along the $z$-axis, say $z=r$, then we find the
  hyperboloid of two sheets $x^2 + y^2= w^2 - r^2$ oriented along the
  $w$-axis.
} let us examine this problem in two steps, each of which can be
visualized in either two or three dimensions.  First, consider the plane
that contains the \hbox{$w$-axis} and the vector ${\bf r}$, and let
$\theta_1$ be the angle between the $w$-axis and the vector ${\bf r}$
in this plane, as shown in Fig.~\hbox{\ref{fig:coord3}a}.  We now
project ${\bf r}$ onto the $w=0$ hyperplane (a three dimensional slice
of four-space), calling the projected vector ${\bf r}_w$. Since the
three-plane $w=0$ lies perpendicular to each of the axes $x$, $y$, and
$z$, the vector ${\bf r}_w$ lies in the three dimensional space shown
in Fig.~\ref{fig:coord3}b, and its length is $ \vert {\bf r}_w \vert =
r\sin\theta_1$.  Let the angle $\theta_2$ be the polar angle between
the $z$-axis and the vector ${\bf r}_w$, while $\theta_3$ is the usual
azimuthal angle $\phi$, as illustrated in \hbox{
Fig.~\ref{fig:coord3}b}. As we vary the three angles and the radial
coordinate, we sweep out a \hbox{four-dimensional} cube (or an
approximate cube) with sides of length $dR_1=r \,d\theta_1$, $dR_2=r
\sin\theta_1 d\theta_2$, $dR_3=r \sin\theta_1\sin\theta_2
\,d\theta_3$, and $dR_4=dr$. This gives a four dimensional volume
element
\begin{eqnarray}
  d^4x 
  \equiv 
  dR_1\,dR_2\,dR_3\,dR_4
  = 
  \sin^2\theta_1 d\theta_1~
  \sin\theta_2 d\theta_2 ~d\theta_3 ~
  r^3\,dr \ ,
\end{eqnarray}
where $0 \le \theta_\ell \le \pi$ for $\ell=1,2$ and $0 \le \theta_3
<2\pi$. As a useful exercise, we can find the four dimensional
hypervolume enclosed by a three-sphere of radius $r$ by integrating
the volume element over the appropriate bounds,
\begin{eqnarray}
  B_4
  =
  \int_0^\pi \!d\theta_1 \sin^2\theta_1\, 
  \int_0^\pi \! d\theta_2\sin\theta_2\, 
  \int_0^{2\pi} \!\! d\theta_3
  \,\int_0^r \! dr^\prime\,  r^{\prime\, 3}
  =
  \frac{1}{2}\, \pi^2\, r^4 \ .
\end{eqnarray}
The derivative of $B_4$ with respect to $r$ gives the hypersurface
area of the \hbox{three-sphere},
\begin{eqnarray}
  S_3
  =
  \frac{dB_4}{dr}
  =
  2\pi^2\, r^3 \ .
\end{eqnarray}
This is analogous to a three dimensional ball of radius $r$ and
volume $B_3=4 \pi\, r^3/3$ bounded by the two-sphere of area $S_2=4\pi
r^2$.

We can readily generalize this procedure to an arbitrary number of
dimensions. Consider a point ${\bf r} \in \mathbb{R}^\nu$ given by the
rectilinear coordinates ${\bf r}=(x_1, x_2, \cdots, x_\nu)$. Let
$\theta_1$ be the angle between the vector ${\bf r}$ and the
$x_1$-axis, in a manner similar to that of Figs.~\ref{fig:coord1}
and~\hbox{\ref{fig:coord3}a}. Note that $dR_1=r d\theta_1$ is the arc
length swept out by ${\bf r}$ as the angle $\theta_1$ is incremented
by $d\theta_1$.  Let us now project ${\bf r}$ onto the hyperplane
$x_1=0$, the $(\nu-1)$-plane normal to the $x_1$-axis and passing
through the origin, calling this projection ${\bf r}_1$: that is to
say, let $ {\bf r} \to {\bf r}_1=(0,x_2,\cdots,x_\nu)$.  The length of
this vector is $r_1 = r \sin \theta_1$.  Let us proceed to the next
step and define the angle $\theta_2$ as the angle between the
$x_2$-axis and ${\bf r}_1$, in which case, as the angle $\theta_2$ is
varied by $d\theta_2$, the vector ${\bf r}_1$ sweeps out an arc of
length $d R_2 = r_1 d\theta_2= r \sin\theta_1\, d\theta_2$. In a
similar fashion, project ${\bf r}_1$ onto the \hbox{$x_2$-plane}, that
is, the plane described by $x_1=0$ and $x_2=0$. This projection is
given by ${\bf r} \to {\bf r}_2=(0,0,x_3,\cdots,x_\nu)$, and the
length of the projection is $r_2=r_1 \sin\theta_2= r \sin\theta_1
\sin\theta_2$.\footnote{\footnoteskip
  Just for good measure, let us do one more iteration. Let $\theta_3$
  be the angle between the $x_3$-axis and the previous projection
  ${\bf r}_2$, in which case $d R_3 = r_2 d\theta_3 = r \sin\theta_1\,
  \sin\theta_2\, d\theta_3$. Let us now project ${\bf r}_2$ onto the
  $x_3$-plane described by $x_1=0$, $x_2=0$, and $x_3=0$, {\em i.e.}
  ${\bf r} \to {\bf r}_3=(0,0,0,x_4,\cdots,x_\nu)$. The length of this
  vector is $r_3=r_2 \sin\theta_3= r \sin\theta_1 \sin\theta_2\,
  \sin\theta_3$, and the next iteration can begin.
} For the general $\ell^{\rm th}$ iteration, let $\theta_\ell$ be the
angle between the $x_\ell$-axis and ${\bf r}_{\ell-1}$, so that $d
R_\ell = r_{\ell-1}\, d\theta_\ell = r \sin\theta_1\, \sin\theta_2\,
\cdots \sin\theta_{\ell-1}\, d\theta_\ell$. In summary, we define
the quantities
\begin{eqnarray}
  && \theta_\ell = \text{angle between the}~{x_\ell}\text{-axis and} 
  ~{\bf r}_{\ell-1}\\
  && dR_\ell = r_{\ell-1}\, d\theta_\ell = r \sin\theta_1\, \sin\theta_2\, 
  \cdots \sin\theta_{\ell-1}\, d\theta_\ell \\
  && {\bf r} \to {\bf r}_\ell = (0,\cdots,0,x_{\ell+1},\cdots, x_\nu) \\
  && r_\ell = r_{\ell-1} \sin\theta_\ell = r \sin\theta_1\, \sin\theta_2\, 
  \cdots   \sin\theta_{\ell-1}\sin\theta_\ell \ ,
\end{eqnarray}
where we have used the fact that $r_{\ell-1} = r \sin\theta_1
\sin\theta_2 \cdots \sin\theta_{\ell-1}$. The last two are lines
provide the projection for the $(\ell+1)^{\rm st}$ step. This gives
the \hbox{$\nu$-dimensional} volume element
\begin{eqnarray}
  d^\nu x 
  =
  \prod_{\ell=1}^\nu dR_\ell
  =
  \sin^{\nu-2}\theta_1 d\theta_1 \cdot 
  \sin^{\nu-3}\theta_2 d\theta_2 \,\cdots\,
  \sin\theta_{\nu-2} d\theta_{\nu-2} \cdot d\theta_{\nu-1}
  \cdot  r^{\nu-1} dr \ .
\label{dnux}
\end{eqnarray}
For notational convenience, I will write the angular measure in
(\ref{dnux}) as $d\Omega_{\nu-1}$, so that
\begin{eqnarray}
  d^\nu x 
  &=& 
  d\Omega_{\nu-1}\, r^{\nu-1} dr \ .
\label{Omegafactor}
\end{eqnarray}
As we proved in the Lecture~I, the integration over all angles gives
the total solid angle
\begin{eqnarray}
  \Omega_{\nu-1} \equiv \int \!d\Omega_{\nu-1} 
  =
  \frac{2\pi^{\nu/2}}{\Gamma(\nu/2)} \ ,
\label{OmegaMinusOne}
\end{eqnarray}
and Table~\ref{table:Omega} illustrates the numerical values of this
solid angle over a wide range of dimensions. Note that $\Omega_{
\nu-1}$ reaches a maximum around $\nu=7$ and then slowly decreases.
\begin{table}[t]
\caption{\footnotesize 
  Solid angle $\Omega_{\nu-1}$ as a function of dimension $\nu$.
\label{table:Omega}
}
\begin{ruledtabular}
\begin{tabular}{lcccccccccc}
  ~$\nu$~ & ~2~ & ~3~ & ~4~ & ~5~ & ~6~ & 7 &  8 & $\cdots$ & 20
\\\hline
   ~$\Omega_{\nu-1}$~ & ~$2\pi$~ & ~$4\pi$~ & ~$2\pi^2$~ & 
  ~$8 \pi^2/3$~ & ~$\pi^3$~ & ~$16 \pi^3/15$~ & ~$\pi^4/3$~ & 
   $\cdots$ & $\pi^{10}/181440$
\\
   ~value~ & ~6.28~ & ~12.6~ & ~19.7~ & ~26.3~ & 
  ~31.0~ & ~33.1~ & ~32.5~ & $\cdots$ 
  & 0.516
\end{tabular} 
\end{ruledtabular}
\end{table}

As a matter of completeness, let us prove (\ref{OmegaMinusOne}) here.
First, consider the one-dimensional Gaussian integral
\begin{eqnarray}
  \int_{-\infty}^\infty dx\, e^{-x^2} = \sqrt{\pi} \ .
\end{eqnarray}
If we multiply both sides together $\nu$ times (with $\nu \in {\mathbb
Z}^+$), we find
\begin{eqnarray}
  (\sqrt{\pi}\,)^\nu 
  = 
  \int_{-\infty}^\infty \!\!dx_1\, e^{-x_1^2} 
  \int_{-\infty}^\infty \!\!dx_2\, e^{-x_2^2} 
  \, \cdots
  \int_{-\infty}^\infty \!\!dx_\nu\, e^{-x_\nu^2} 
  =
  \int \! d^\nu x\, e^{-{\bf r}^2} \ ,
\label{gaussnu}
\end{eqnarray}
where the vector ${\bf r}$ in the exponential of the last expression
is the $\nu$-dimensional vector ${\bf r}=(x_1, x_2, \cdots, x_\nu)$,
and ${\bf r}^2 = {\bf r} \cdot {\bf r} =\sum_{\ell=1}^\nu
x_\ell^2$\,. As in (\ref{Omegafactor}), we can factor the angular
integrals out of the right-hand-side of (\ref{gaussnu}), and the
remaining one-dimensional integral can be converted to a Gamma
function with the change of variables $t=r^2$\,:
\begin{eqnarray}
  \pi^{\nu/2}
  = \!\!
  \int \!\! d\Omega_{\nu-1} \cdot \!\!
  \int_0^\infty \!\!\! dr\, r^{\nu-1}\,e^{-r^2} 
  \! = \!\!
  \int \!\! d\Omega_{\nu-1} \cdot \frac{1}{2}\, \Gamma(\nu/2) \,.
\label{pinuomega}
\end{eqnarray}
Solving for $\int\!d\Omega_{\nu-1}$ in (\ref{pinuomega}) gives
(\ref{OmegaMinusOne}).

In calculating the temperature equilibration rate and the charged
particle stopping power, we encounter integrals of the form
\begin{eqnarray}
  I_1(\nu) 
  &\equiv& 
  \int d^\nu x ~ f_1(r) ~\,
  = ~
  \Omega_{\nu-1} \int_0^\infty \! dr\, r^{\nu-1}\, f_1(r)
\label{fonedr}
\\[5pt]
  I_2(\nu) 
  &\equiv& 
  \int \! d^\nu x\, f_2(r,\theta) 
  = ~
  \Omega_{\nu-2} \int_0^\infty \! dr\, r^{\nu-1}
  \int_0^\pi \!d\theta\,\sin^{\nu-2}\theta\,f_2(r,\theta) \ ,
\label{ftwodrdth}
\end{eqnarray}
respectively, with $\nu \in \mathbb{Z}^+$. The exact forms of $f_1$
and $f_2$ do not concern us here, except that their angular dependence
is determined by the following considerations: the integral
(\ref{fonedr}) is spherically symmetric since the energy exchange
between plasma species is isotropic, while in the latter integral
(\ref{ftwodrdth}), the motion of the charged particle defines a
preferred direction around which one must integrate. The integrals
$I_1$ and $I_2$ can be viewed as functions defined on the positive
integers, and as discussed at length in Lecture~I~\cite{bps1},
Carlson's Theorem~\cite{Carlsonth} ensures that there is a unique
analytic continuation onto the complex plane.  As our first
application in this section, let us see how the expressions
(\ref{fonedr}) and (\ref{ftwodrdth}) provide a means by which to
easily and conveniently perform this analytic continuation to complex
values of the spatial dimension $\nu$, thereby rendering $\nu$ truly
arbitrary. First, the solid angles $\Omega_{\nu-1}$ and
$\Omega_{\nu-2}$ are composed of a simple exponential factor
$\pi^{\nu/2}$ and a Gamma function, whose analytic continuations have
been well studied. As for the integrals, simply treat $\nu$ as a
complex parameter, performing the one dimensional integral
(\ref{fonedr}) and the double integral (\ref{ftwodrdth}) in the usual
manner of ordinary calculus.  This provides functions $I_1(\nu)$ and
$I_2(\nu)$ of a complex argument \hbox{$\nu \in \mathbb{C}$}, in
fulfillment of Carlson's Theorem. Double integrals of the form
(\ref{ftwodrdth}) were used extensively in Ref.~\cite{bps} to
calculate the stopping power, where the angle $\theta$ is determined
by the direction of motion of the charged particle. Calculating the
temperature equilibration rate, on the other hand, requires only the
simpler one dimensional integral (\ref{fonedr}), as the energy
exchange in this process is isotropic.

\subsubsection{\label{sec:areas} 
The Hypervolume of Spheres, Disks, and Cylinders}

\begin{figure}[t]
\includegraphics[scale=0.35]{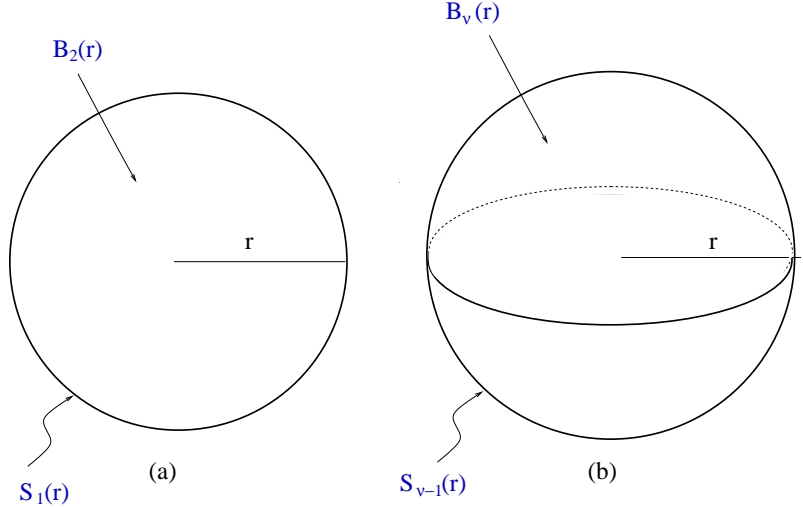}
\vskip-0.3cm 
\caption{\captionskip
  A $(\nu-1)$-dimensional sphere $S_{\nu-1}$ of radius $r$ bounds the
  $\nu$-dimensional ball $B_\nu(r)$ of radius $r$. By integrating over
  successive shells of area, we can find the volume by $B_\nu(r) =
  \int_0^r dr^\prime S_{\nu-1}(r^\prime)$; or conversely
  $S_\nu(r)=B_\nu^\prime(r)$.
}
\label{fig:sphere}
\end{figure}

We shall now calculate the hypervolume of several useful geometric
objects. Let us first consider a \hbox{$\nu$-dimensional} ball of
radius $r$, defined by the set of points \hbox{ ${\bf x} \in
\mathbb{R}^\nu$} for which $\vert {\bf x} \vert \le r$. We will denote
this object by $B_\nu(r)$, and in two and three dimensions this is a
disk and a spherical, both volume centered at the origin.  We can find
the $\nu$-dimensional hypervolume of the ball $B_\nu(r)$ by simply
integrating (\ref{dnux}) over all permissible values of the
coordinates. It should cause no confusion to denote the hypervolume of
the region $B_\nu(r)$ by the same symbol, and using
(\ref{OmegaMinusOne}) we find
\begin{eqnarray}
  B_\nu(r)
  &=&
  \int d\Omega_{\nu-1}
  \int_0^r dr^\prime \, r^{\prime\,\nu-1} 
  =
  \frac{\pi^{\nu/2}}{\Gamma(\nu/2+1)}\,r^\nu \ .
\label{Bnu}
\end{eqnarray}
The boundary of $B_\nu(r)$ is a $(\nu\!-\!1)$-dimensional sphere
$S_{\nu-1}(r)$ defined by $\vert {\bf x} \vert = r$, or
\hbox{$\sum_{\ell=1}^\nu x_\ell^2 = r^2$}.  By differentiating (\ref{Bnu})
with respect to the radius $r$, we can also find the hyperarea of a
\hbox{ $(\nu\!  -\!  1)$-dimensional} sphere $S_{\nu-1}(r)$ of radius
$r$ in ${\mathbb R}^\nu$,
\begin{eqnarray}
  S_{\nu-1}(r)   
  &=&
  \frac{d B_\nu}{dr}
  =
  \frac{2\pi^{\nu/2}}{\Gamma(\nu/2)}\, r^{\nu-1} 
  =
  \Omega_{\nu-1}\,r^{\nu-1}\ .
\label{Snu}
\end{eqnarray}
For brevity, I have denoted the hyperarea by the same symbol
$S_{\nu-1}(r)$ as the sphere itself, which is simply the \hbox{
$(\nu\!-\!1)$-dimensional} boundary of the region $B_\nu(r)$. This is
illustrated in Fig.~\ref{fig:sphere}.  The distinction I am making
between ``hypervolume'' and ``hyperarea'' is somewhat arbitrary, since
these are both terms involving regions in a higher dimensional
space. When I wish to talk about a \hbox{ $\nu$-dimensional} subregion
of the hyperspace $\mathbb{R}^\nu$, such as $B_\nu(r)$, I will use the
term hypervolume. On the other hand, when I wish to emphasize a
boundary region of a hypervolume, such as $S_{\nu-1}(r)$, I will use
the term ``hyperarea.'' Regarding the usage of the term ``solid
angle,'' suppose we keep the radius $r$ fixed but vary the angles
$\theta_i$ over ranges $d\theta_i$. The region swept out by this
procedure lies on the $(\nu\!-\!1)$-dimensional sphere $S_{\nu-1}(r)$
with a hyperarea $dS_{\nu-1}= d\Omega_{\nu-1}\, r^{\nu-1}$. We are
therefore justified in calling $d\Omega_{\nu-1}$ the solid angle in
$\nu$ dimensions.

\begin{figure}[t]
\includegraphics[scale=0.3]{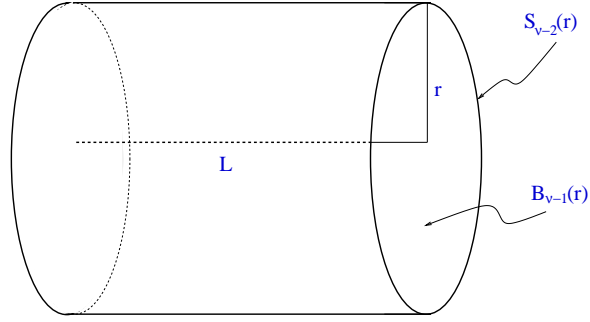}
\caption{\captionskip 
  The hyperarea of a hypercylinder $C_{\nu-1}(r,L)$ of length $L$ and
  radius $r$ is $C_{\nu-1}(r,L) = S_{\nu-2}(r)\cdot L$, and the hypervolume
  bounded by the cylinder is $V_\nu(r,L)= B_{\nu-1}(r)\cdot L$.
}
\label{fig:cylinder}
\end{figure}

Finally, let us discuss the $(\nu\!-\!1)$-dimensional cylindrical
$C_{\nu-1}(r,L)$ of radius $r$ and length $L$. Again, it is easiest to
argue from analogy in three dimensions.  To form a two-cylinder
$C_2(r,L)$ in $\mathbb{R}^3$, we let a two dimensional disk $B_2(r)$
sweep out a volume as it moves a distance $L$ in the orthogonal
direction, which is illustrated in Fig.~\ref{fig:cylinder}a.
Similarly, a corresponding \hbox{ $(\nu\!-\!1)$-dimensional} cylinder
is formed by letting a \hbox{$(\nu\!-\!1)$-dimensional} ball
$B_{\nu-1}(r)$ sweep out a distance $L$ along the orthogonal axis, as
illustrated in Fig.~\ref{fig:cylinder}b.  Therefore, the hyperarea of
the \hbox{ $(\nu\!-\!1)$-dimensional} cylinder is
\begin{eqnarray}
  C_{\nu-1}(r,L) 
  &=&
  S_{\nu-2}(r) \cdot L  \ .
\label{Cnu}
\end{eqnarray}
The $\nu$-dimensional hypervolume enclosed by this cylinder is
\begin{eqnarray}
  V_\nu(r,L) 
  &=&
  B_{\nu-1}(r) \cdot L  \ .
\label{BCnu}
\end{eqnarray}

\pagebreak
\subsection{The Cross Section}

\begin{figure}[t]
\includegraphics[scale=0.45]{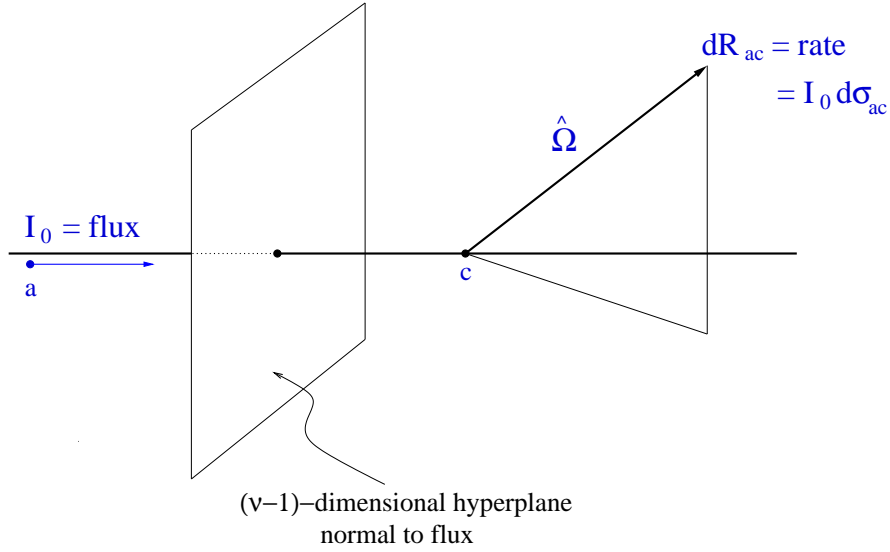}
\caption{\captionskip 
  Definition of the cross section in a general number of
  dimensions. The incident flux $I_0$ of species $a$ is the rate of
  particles per unit hyperarea normal to the flow. The units of $I_0$
  are $\text{L}^{1-\nu} \cdot \text{T}^{-1}$, where $\text{L}$ and
  $\text{T}$ denote the units of space and time. By definition, 
  the differential {\em cross section} $d\sigma_{ac}$ is related 
  to the rate $d R_{ac}$, each at angular position $\hat\Omega$, 
  by 
  $d R_{ac}(\hat\Omega)= I_0 \, d \sigma_{ac}(\hat\Omega)$. The cross
  section per unit solid angle about the direction $\hat\Omega$ is
  denoted by $d\sigma_{ac}/d\Omega$. Except for the specification of
  $\hat\Omega$, this definition does not depend upon the
  dimensionality of space, and the units of $d\sigma_{ac}$ are
  $\text{L}^{\nu-1}$.
}
\label{fig:cross}
\end{figure}

As a physical application of hyperspherical coordinates, let us
calculate the form of the classical ``cross section'' in
$\nu$-dimensions. For simplicity we will consider a projectile
striking a {\em fixed} target, although we can perform a similar
analysis in the center-of-mass frame of the two particles.  Such a
scattering experiment is illustrated in Fig.~\ref{fig:cross}, in which
a beam of incident particles, denoted by the label $a$, is fired at a
target $c$ with incident flux $I_0$. The rate $dR_{ac}(\hat\Omega)$ at
which the scattered $a$-particles enter a given solid angle
$d\Omega_{\nu-1}$ about the direction $\hat{\Omega}$ is then measured.
The flux $I_0$ is a characterization of the rate at which particles
move along the beam axis. In $\nu$ dimensions, the spatial region
normal to the axis is a $(\nu\!-\!1)$-dimensional hyperplane, and the
flux $I_0$ is the number of particles per second per unit hyperarea
passing through this plane.  For example, if the beam direction is
$\hat {\bf n}$, then the number of particles in a time interval $dt$
passing through a hyperarea $dA_{\hat{\bf n}}$ normal to $\hat{\bf n}$
is given by $dN = I_0 \cdot dA_{\hat{\bf n}}\cdot dt$. The engineering
units of $I_0$ are therefore $L^{1-\nu} \cdot T^{-1}$. In analogy with
the usual cross section in three dimensions, we define $d \sigma_{ac}$
through
\begin{equation}
  d \sigma_{ac} \cdot I_0 = dR_{ac} \ ,
\end{equation}
and $d \sigma_{ac}$ therefore has engineering units of $L^{\nu-1}$.

Suppose the scattering center is a central force, such as the
\hbox{$\nu$-dimensional} Coulomb potential. The particle is confined
to a two-dimensional plane for central potential motion, and this
holds true even in $\nu$ dimensions.  Let $b$ denote the impact
parameter of projectile.  As the particle traverses its plane of
motion, its position is uniquely characterized by a function
$b=b(\theta)$, where $\theta$ is the angle between the beam direction
and the projectile (with the scattering center defining the
origin). From Fig.~\ref{fig:cylinder}, the number of particles per
unit time passing through the hyperannulus of width $db$ and radius
$b$ is is $dN= S_{\nu-2}(b) db \cdot I_0 = \Omega_{\nu-2}\, b^{\nu-2}
\, db \cdot I_0$, and by particle number conservation, the same number
of scattered particles reaches the hyperannulus at $\theta$. The cross
section in a $\nu$-dimensional central potential is therefore given by
\begin{equation}
  d \sigma_{ac} =\Omega_{\nu -2} \, b^{\nu -2} \,db \ .
\end{equation}
This is Eq.~(8.31) of Ref.~\cite{bps}, the starting point for the
classical calculation.  The cross section will appear in the Boltzmann
equation. To include two-body quantum scattering effects, we replace
the classical cross section by the quantum cross section:
\begin{eqnarray}
  \vert{\bf v}_a - {\bf v}_c \vert\,
  d\sigma_{ac}
  =
  \big\vert\, T \,\big\vert^2\, 
  \frac{d^\nu p_c}{(2\pi\hbar)^\nu}\,
  \frac{d^\nu p_a}{(2\pi\hbar)^\nu} \ ,
\label{dSigmaa}
\end{eqnarray}
where $T$ is the quantum scattering amplitude. In the calculations
that follow, we shall use work in the extreme quantum limit where
the Born approximation for the amplitude can be employed.

\subsection{\label{sec:physdim} The Coulomb Potential in Arbitrary Dimensions}

Now that we have discussed the cross section in an arbitrary central
potential, let us concentrate on the special case of the Coulomb
potential.  The physics of dimensional continuation is contained in
the \hbox{$\nu$-dependence} of the Coulomb potential in
\hbox{$\nu$-dimensional space}, which ensures that short distance
physics is emphasized in $\nu>3$ and long distance physics in
$\nu<3$. Changing the spatial dimension about $\nu=3$ therefore acts
as a ``physics sieve.'' Let us first construct the electric field of a
point charge in $\nu$ dimensions. Maxwell's equations are easily
generalized to an arbitrary number of dimensions, and in particular, 
we can write 
\begin{eqnarray}
  {\bm\nabla}\!\cdot\!{\bf E}({\bf x}) 
  = \rho({\bf x}) \ ,
\label{Gaussdiff}
\end{eqnarray}
where ${\bf E}=(E_1, \cdots, E_\nu)$ is the electric field vector and
${\bm\nabla}=(\partial/\partial x_1, \cdots , \partial/\partial
x_\nu)$ is the \hbox{$\nu$-dimensional} spatial gradient. The charge
density $\rho$ has engineering units of charge divided length to the
$\nu^{\rm th}$ power, which I will write as $Q/L^\nu$. In integral
form, the equation can be written
\begin{eqnarray}
  \int_\Sigma d^\nu x \,
  {\bm\nabla}\!\cdot\!{\bf E} = e \ ,
\label{Gaussint}
\end{eqnarray}
where $e$ is the total electric charge contained in the hypervolume
$\Sigma$. Note that the dimensionality of space is now explicitly
indicated by the integration measure. We can employ the usual symmetry
argument to find the electric field of a point source at the
origin. Let $B_r$ be the $\nu$-dimensional ball of radius $r$ centered
on the point charge $e$, and denote the $(\nu\!-\!1)$-dimensional
hyperspherical boundary by $S_r$. By symmetry, the field ${\bf E}(
{\bf x})$ points radially outward with a magnitude $E(r)$ along the
direction $\hat{\bf x}$ normal to $S_r$. The length $E(r)$ depends
only upon the radial distance $r = \vert {\bf x} \vert$ and not upon
its angular location along $S_r$. The divergence theorem holds in an
arbitrary number of dimensions, and since the hyperarea of $S_r$ is
given by (\ref{Snu}), we find:
\begin{eqnarray}
  e
  = 
  \int_{B_r} d^\nu x \,{\bm\nabla}\!\cdot\!{\bf E}
  = 
  \oint_{S_r}\! d{\bf A}\!\cdot\! {\bf E}
  =
  \Omega_{\nu-1}\,r^{\nu-1} \cdot  E(r) 
  ~~~~~ \text{with}~~~\Omega_{\nu-1} = 
  \frac{2\pi^{\nu/2}}{\Gamma(\nu/2)} \ .
\label{eintGamma}
\end{eqnarray}
The electric field is therefore given by 
\begin{eqnarray}
  {\bf E}({\bf x}) 
  = 
  \frac{\Gamma(\nu/2)}{2 \pi^{\nu/2}}\,
  \frac{e}{r^{\nu-1}}\,\hat{\bf x} \ ,
\label{Enu}
\end{eqnarray}
where we are using the notation ${\bf x}= r\, \hat{\bf x}$, with
$\hat{\bf x}$ being a unit vector pointing in the direction of
${\bf x}$.

\begin{figure}[t]
\includegraphics[scale=0.45]{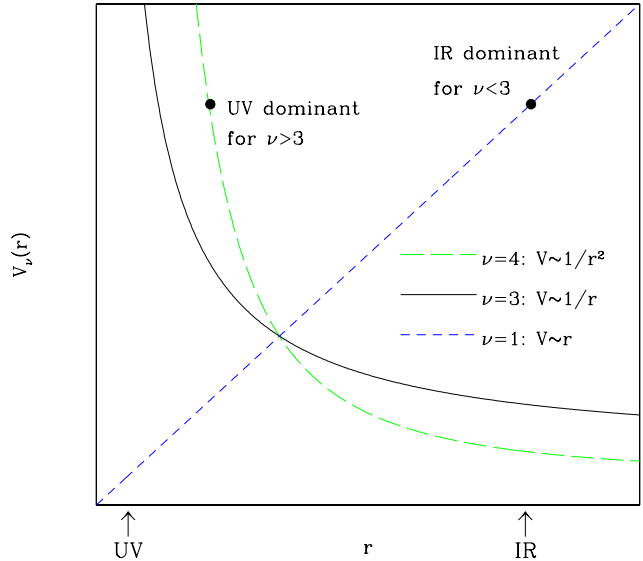}
\vskip-0.8cm 
\caption{\captionskip
  Short-distance ultraviolet (UV) physics dominates in dimensions
  $\nu>3$. Long-distance or infrared (IR) physics dominates when
  $\nu<3$. UV and IR physics are equally important in $\nu=3$.  }
\label{fig:coulomb}
\end{figure}

I find it more convenient to work with the electric
potential, a scalar quantity $\phi(r)$ defined by $E(r) = -
d\phi(r)/dr$. In fact, I will work with the potential energy
\hbox{$V_\nu = e\, \phi(r)$}, so that
\begin{eqnarray}
  V_\nu({\bf x})
  = 
  \frac{\Gamma(\nu/2-1)}{4 \pi^{\nu/2}} \,
  \frac{e^2}{r^{\nu-2}} \ ,
\label{Vnu}
\end{eqnarray}
where I have appended a subscript to the potential energy to remind us
that we are working in $\nu$ dimensions. For two charges $e_a$ and
$e_b$ separated by a distance $r$, one only need replace $e^2$ by the
product $e_a\,e_b$. For $\nu\!=\!3$, the geometric factor in
(\ref{Vnu}) becomes $1/4\pi$, which is the origin of the $4\pi$ of
rationalized units.  Figure~\ref{fig:coulomb} shows the Coulomb
potential for $\nu\!=\!3$, along with two representative dimensions on
either side of $\nu\!=\!3$.  As the figure illustrates, the short
distance behavior of the Coulomb potential becomes more pronounced in
higher dimensions, while long-distances are emphasized in lower
dimensions. For aesthetic reasons, the arbitrary integration constant
for the potential energy has been adjusted in each case so that all
three graphs intersect at a single point. This figure illustrates
quite dramatically that by simply dialing the dimension $\nu$, we can
dial a potential $V_\nu(r)$ that filters either long-distance or
short-distance physics.

In the Born approximation to quantum Coulomb scattering, which we
shall employ shortly, we need the Fourier transform of the Coulomb
potential (\ref{Vnu}). Unlike the spatial representation $V_\nu({\bf
x})$, the Fourier representation of the $\nu$-dimensional Coulomb
potential takes the same form in any dimension, namely,
\begin{eqnarray}
  \tilde V_\nu({\bf k})
  =
  - \frac{e^2}{k^2} \ ,
\label{CoulombXform}
\end{eqnarray}
where $k^2=\sum_{\ell=1}^\nu k_\ell^2$ is just the square of the norm
of the $\nu$-dimensional wave number ~${\bf k}$, and I am using the
conventions
\begin{eqnarray}
  V_\nu({\bf x}) 
  &=& 
  \int \!\frac{d^\nu k}{(2\pi)^\nu} \,
  e^{-i{\bf x}\cdot{\bf k}}\, \tilde V_\nu({\bf k})
\label{VxVtilde}
\\[5pt]
  \tilde V_\nu({\bf k}) 
  &=& 
  \int \! d^\nu x \,e^{i{\bf x}\cdot{\bf k}}\, 
  V_\nu({\bf x}) \ .
\label{VtildeVx} 
\end{eqnarray}
With these conventions, the amplitude in the Born approximation 
in any dimension is given by 
\begin{eqnarray}
  T_\smB 
  = 
  \hbar \, \frac{e^2}{q^2} \ ,
\label{TsmB}
\end{eqnarray}
where ${\bf q}={\bf p}_a - {\bf p}_b$ is the momentum transfer
during the collision. This is a function only of the square of
its argument $q^2$. In particular, the Born approximation does
not introduce dependence upon the center-of-momentum energy $W$,
and this is what renders its use so convenient.

Expression (\ref{CoulombXform}) for the Fourier transform of the
potential (\ref{Vnu}) can be established in a number of ways, perhaps
the easiest being an straightforward application of Laplace's
equation,
\begin{eqnarray}
  \nabla^2 V_\nu({\bf x}) 
  = e^2\, \delta^{(\nu)}({\bf x}) \ .
\label{laplaceV}
\end{eqnarray}
Upon inserting (\ref{VxVtilde}) for $V_\nu({\bf x})$ into
(\ref{laplaceV}) and using the integral representation of the
delta-function, we can write Laplace's equation in the form
\begin{eqnarray}
 -\int \!
  \frac{d^\nu k}{(2\pi)^\nu} \,
  e^{-i{\bf x}\cdot{\bf k}}\, k^2\, \tilde V_\nu({\bf k})
  =
  e^2 \int \!
  \frac{d^\nu k}{(2\pi)^\nu} \, e^{-i{\bf x}\cdot{\bf k}} \ ,
\end{eqnarray}
and solving for $\tilde V({\bf k})$ provides (\ref{CoulombXform}). 
It might also be informative to prove (\ref{CoulombXform}) using the
more direct approach of performing the Fourier transform directly.
Substituting the Coulomb potential (\ref{Vnu}) into (\ref{VtildeVx}),
and then using (\ref{ftwodrdth}) to rewrite the $\nu$-dimensional 
integral, we find
\begin{eqnarray}
  \tilde V_\nu({\bf k}) 
  &=&  \Omega_{\nu-2} \int_0^\infty \! dr\, r^{\nu-1}
  \int_0^\pi \!d\theta\,\sin^{\nu-2}\theta\,e^{i r k \cos\theta}
  \cdot 
  \frac{\Gamma(\nu/2-1)}{4 \pi^{\nu/2}} \,
  \frac{e^2}{r^{\nu-2}} 
\\[10pt]
  &=&
  \frac{e^2}{2\sqrt{\pi}}\,
  \frac{\Gamma(\nu/2-1)}{\Gamma(\nu/2-1/2)}\,
  \int_0^\infty \!\!   r dr \int_0^1\! du\,
  (1-u^2)^{(\nu-3)/2}\,   \Big[ e^{i k\, r u} + 
  e^{-i k\, r u} \Big]\ ,
\label{tildeVtwo}
\end{eqnarray}
where we have made the change of variables $u=\cos\theta$. It is
convenient to keep the exponential terms in square brackets rather
than converting their sum into a cosine term.  We will perform the 
$r$-integration by deforming the contour slightly off the real axis, 
\begin{eqnarray}
  \int_0^{\infty} \!\!  
  dr\, r\, \Big[e^{i\,k\, r u} + e^{-i\,k\, r u} \Big]
  &=&
  \int_0^{\infty+i\epsilon} \!\!  dr\, r\, e^{i\,k\, r u} 
  +
  \int_0^{\infty-i\epsilon} \!\!  dr\, r\, e^{-i\,k\, r u} 
\\
  &=&
  -\frac{2}{(k u)^2} \ .
\end{eqnarray}
Upon substituting this back into (\ref{tildeVtwo}) and changing
variables to $t=u^2$ we can write
\begin{eqnarray}
  \tilde V_\nu({\bf k}) 
  &=&
  \frac{e^2}{2\sqrt{\pi}}\,\frac{1}{k^2}\,
  \frac{\Gamma(\nu/2-1)}{\Gamma(\nu/2-1/2)}\,
  \int_0^1\! du\,(1-t)^{(\nu-3)/2}\,   t^{-3/2} \ ,
\label{tildeVthree}
\end{eqnarray}
where the second term in the integrand introduces the pole $1/(\nu-3)$
into physical quantities, and the $t$-integral takes the form of the
Euler Beta function
\begin{eqnarray}
  B(x,y) 
  = 
  \int_0^1 \! dt\, t^{x-1}\, (1-t)^{y-1}
  =
  \frac{\Gamma(x) \Gamma(y)}{\Gamma(x+y)} 
\label{Bxy}
\end{eqnarray}
with $x=-1/2$ and $y=\nu/2-1$. Using $\Gamma(-1/2)=-2\sqrt{\pi}$
gives (\ref{CoulombXform}). In Section~\ref{sec:LBE} we
will need yet another representation of the Beta function, which
I record here for convenience: 
\begin{eqnarray}
  B(x,y)=\int_0^\infty dt\, t^{x-1} (1+t)^{-x-y}= 
  \frac{\Gamma(x)\Gamma(y)}{\Gamma(x+y)} \ .
\label{Bxyanother}
\end{eqnarray}

\subsection{\label{sec:keqnu} Kinetic Equations in Arbitrary Dimensions}

\subsubsection{\label{sec:distf} Distribution Functions}

A particle in a space $\mathbb{R}^\nu$ of arbitrary dimension $\nu \in
\mathbb{Z}^+$ is fully characterized by its position and momentum
${\bf x}$ and ${\bf p}$, which have rectilinear coordinates $x_\ell$
and $p_\ell$ for $\ell=1, \cdots, \nu$. \hbox{I will} often denote the
square and the magnitude of the momentum by $p^2 = {\bf p} \cdot {\bf
p} = \sum_{\ell=1}^\nu p_\ell^2$ and $p = \vert {\bf p}\vert$,
respectively. For example, $p^{-3}$ is shorthand for $\vert {\bf p}
\vert^{-3}= \left(\,\sum_{\ell=1}^\nu p_\ell^2 \,\right)^{-3/2}$.  A
swarm of particles distributed over position and momentum values
is characterized by a distribution function $f$ defined by
\begin{eqnarray}
\nonumber 
  f({\bf x},{\bf p},t)\,\frac{d^\nu x\,d^\nu p}{(2\pi\hbar)^\nu}\, 
  &\equiv& 
  \text{number of particles in a hypervolume $d^\nu x$ about ${\bf x}$}
\\[-10pt] && 
  \text{and $d^\nu p$ about ${\bf p}$ 
  at time $t$} \ .
\label{fnu}
\end{eqnarray}
The factor of $(2\pi\hbar)^\nu$ in the denominator is a conventional
normalization factor, and for a spatially uniform distribution
$f_a$ this gives the normalization
\begin{eqnarray}
  \int\! \frac{d^\nu p_a}{(2\pi\hbar)^\nu}\,
  f_a({\bf p}_a) &=& n_a \ ,
\label{fbnorm}
\end{eqnarray}
where $n_a$ is the number density of $a$-type particles. That is to
say, $n_a\,d^\nu x$ is the number of particles of species $a$ in a
hypervolume $d^\nu x$, and the engineering units of $n_a$ are
therefore $L^{-\nu}$. From (\ref{fbnorm}), we see that a normalized
Maxwell-Boltzmann distribution at temperature $T_a$ and number density
of $n_a$ is given by
\begin{eqnarray}
  f_a({\bf p}_a) 
  =
  n_a
  \left(\frac{2\pi\hbar^2 \beta_a}{m_a}\right)^{\nu/2}
  \exp\left\{-\beta_a\, \frac{p_a^2}{2m_a}\right\}  
  =
  n_a\, \lambda_a^\nu\, e^{-\beta_a E_a} \ ,
\label{defFiA}
\end{eqnarray}
where $E_a = p_a^2/2 m_a$ is the kinetic energy and $\beta_a=1/T_a$ is
the inverse temperature in energy units. The thermal wave length for
species $a$ is defined by
\begin{eqnarray}
  \lambda_a
  &=& 
  \hbar \left( \frac{2\pi \beta_a}{m_a}\right)^{1/2}  \ .
\label{deflambd}
\end{eqnarray}
Consequently, the spatial density of the kinetic energy of species $a$
is given by
\begin{eqnarray}
  {\cal E}_a
  = 
  \int\!\frac{d^\nu p_a}{(2\pi\hbar)^\nu} \,
  \frac{p_a^2 }{2 m_a}~ f_a({\bf p}_a,t) \ ,
\label{Ea}
\end{eqnarray}
where $f_a$ is the corresponding distribution function.

Suppose now that the rate of change in the distribution function $f_a$
is determined by some kinetic equation
\begin{eqnarray}
  \frac{\partial f_a}{\partial t} + 
  {\bf v}_a \! \cdot \!{\bm\nabla}_{\!\!x} f_a
  &=& 
  {\sum}_b K_{ab}[f]   \ ,
\label{kequgen}
\end{eqnarray}
where ${\bm\nabla}_x$ is the $\nu$-dimensional gradient in position
space, ${\bf v}_a={\bf p}_a/m_a$ is the particle velocity, and
$K_{ab}$ is a scattering kernel between particles of type $a$ and type
$b$. When the distribution is spatially uniform we may set the
convective term to zero, ${\bf v}_a \!\cdot\!{\bm\nabla}_{\!x}f_a =
0$, in which case the time rate of change in the kinetic-energy
density (\ref{Ea}) of the $a$-species is given by
\begin{eqnarray}
  \frac{d {\cal E}_a}{dt} 
  = 
  \int\!\frac{d^\nu p_a}{(2\pi\hbar)^\nu} \,
  \frac{p_a^2 }{2 m_a}~
  \frac{\partial f_a}{\partial t}({\bf p}_a,t)
  = 
  {\sum_b}\int \frac{d^\nu p_a}{(2\pi\hbar)^\nu} \,
  \frac{p_a^2 }{2 m_a}~ K_{ab}[f] \ .
\label{dedtbenu}
\end{eqnarray}
We can therefore identify the rate of change in the kinetic-energy
density of species $a$, resulting from its Coulomb interactions with
species $b$, by the expression
\begin{eqnarray}
  \frac{d {\cal E}_{ab}}{dt} 
  = 
  \int \frac{d^\nu p_a}{(2\pi\hbar)^\nu} \,
  \frac{p_a^2 }{2 m_a}~ K_{ab}[f] \ .
\label{dedtbenuab}
\end{eqnarray}
Since we are taking each species to be in thermal equilibrium with
itself, but not necessarily with the other species, each species $b$
is characterized by a unique temperature $T_b$. The rate $d{\cal
E}_{ab}/dt$ is therefore proportional to the temperature difference
between these species, and we write
\begin{eqnarray}
  \frac{d {\cal E}_{ab}}{dt} 
  = 
  - {\cal C}_{ab}\,(T_a - T_b) \ ,
\label{Cabdef}
\end{eqnarray}
where ${\cal C}_{ab}$ is called the rate coefficient. By performing
the integrals in (\ref{dedtbenuab}) exactly, and then comparing with
(\ref{Cabdef}), we may calculate the coefficients ${\cal C}_{ab}$
exactly. If the ions are at a common temperature $T_\smI$, then it is
more convenient to calculate the rate coefficient between the
electrons and the collective set of ions, the coefficient ${\cal C
}_{e\smI}={\sum}_i\, C_{ei}$ of (\ref{dedteI}).

\subsubsection{\label{sec:BEnu} The Boltzmann Equation}

The derivation of the Boltzmann equation presented in Section~3.3 of
Ref.~\cite{huang} goes through unscathed in a general number of
dimensions, and the scattering kernel is completely finite when $\nu >
3$. The derivation breaks down in $\nu \le 3$ (that is to say, for
$\nu=1,2,3$), because in these dimensions the scattering kernel for
the Coulomb interaction diverges. This behavior for the Boltzmann
equation arises because the Coulomb interaction emphasizes the short
distance physics when $\nu>3$, while the scattering kernel of the
Boltzmann equation is designed to capture such short distance physics.
I~will write the Boltzmann equation in schematic form as
\begin{eqnarray}
  \frac{\partial f_a}{\partial t} + 
  {\bf v}_a \! \cdot \!{\bm\nabla}_x f_a 
  &=& 
  {\sum}_b B_{ab}[f]   ~~~~:\, \nu> 3 \ ,
\label{BEsimpnu}
\end{eqnarray}
or in explicit form by writing the full scattering kernel as
\begin{eqnarray}
\nonumber
  B_{ab}[f]
  &=&
  \int \frac{d^\nu p_b}{(2\pi\hbar)^\nu}\,d\Omega\,
  \vert{\bf v}_b - {\bf v}_a \vert\, \frac{d\sigma_{ab}}{d\Omega}\,
  \bigg\{
  f_a({\bf p}_a^\prime) f_b({\bf p}_b^\prime)
  -
  f_a({\bf p}_a) f_b({\bf p}_b)
  \bigg\} 
\\[5pt] && \hskip-0.5cm 
  (2\pi\hbar)^\nu\,\delta^\nu\!\Big( {\bf p}_a^\prime + {\bf p}_b^\prime - 
  {\bf p}_a - {\bf p}_b \Big)\,
  (2\pi\hbar) \delta\Big(E_a^\prime + E_b^\prime - E_a - E_b\Big) \ ,
\label{BEfeSigma}
\end{eqnarray}
with $E_c^\prime=p_c^{\prime\,2}/2m_c$ and $E_c=p_c^{\,2}/2m_c$.  See
Fig.~\ref{fig:cross} for an explanation of the cross section
$d\sigma_{ab}$ in $\nu$ spatial dimensions.  We can include the
quantum effects of two-body scattering, to the order in $g$ to which
we are working, by replacing the classical cross section by the
corresponding quantum cross section defined in (\ref{dSigmaa}).  It
then becomes necessary to calculate the quantum transition amplitude
$T(ab \to a^\prime b^\prime) \equiv T_{a^\prime b^\prime;\, ab}$, and
rewriting (\ref{dSigmaa}) in the form
\begin{eqnarray}
  \vert{\bf v}_b - {\bf v}_a \vert\,
  d\sigma_{ab}
  =
  \big\vert T_{a^\prime b^\prime;\,ab} \big\vert^2\, 
  \frac{d^\nu p_a^\prime}{(2\pi\hbar)^\nu}\,
  \frac{d^\nu p_b^\prime}{(2\pi\hbar)^\nu} \ ,
\label{dSigma}
\end{eqnarray}
one can then include quantum effects by using the scattering kernel
\begin{eqnarray}
\nonumber
  B_{ab}[f]
  &=&
  \int \frac{d^\nu p_a^\prime}{(2\pi\hbar)^\nu}\,
  \frac{d^\nu p_b^\prime}{(2\pi\hbar)^\nu}\,
  \frac{d^\nu p_b}{(2\pi\hbar)^\nu}\,\big\vert T_{a^\prime b^\prime;\,ab} 
  \big\vert^2\, 
  \bigg\{
  f_a({\bf p}_a^\prime) f_b({\bf p}_b^\prime)
  -
  f_a({\bf p}_a) f_b({\bf p}_b)
  \bigg\} 
\\[5pt] && \hskip-0.5cm 
  (2\pi\hbar)^\nu\,\delta^\nu\!\Big( {\bf p}_a^\prime + {\bf p}_b^\prime - 
  {\bf p}_a - {\bf p}_b \Big)\,
  (2\pi\hbar) \delta\Big(E_a^\prime + E_b^\prime - E_a - E_b\Big) \ .
\label{BEfe}
\end{eqnarray}
For simplicity, in Sec.~\ref{sec:BE} we shall use the Born
approximation (\ref{TsmB}) for the transition amplitude, which
corresponds to taking the extreme quantum limit.  When $\nu>3$,
expression (\ref{dedtbenuab}) allows us to write the rate of change of
the energy density resulting from the now finite Boltzmann equation as
\begin{eqnarray}
  \frac{d{\cal E}_{ab}^\smGT}{dt}
  =
  \int \!\frac{d^{\,\nu}p_a}{(2\pi\hbar)^\nu}~\frac{p_a^2}{2 m_a}~
  B_{ab}[f] ~~~~ : ~~ \nu>3 \ .
\label{dedtgtthree}
\end{eqnarray}
I have used a ``greater than'' superscript to remind us that we should
calculate (\ref{dedtgtthree}) in dimensions greater than three.  

\subsubsection{\label{sec:LBEnu} The Lenard-Balescu Equation}

In dimensions less than three one finds a complementary situation to
the Boltzmann equation, namely, the derivation of the Lenard-Balescu
equation is rigorous and completely finite when $\nu < 3$. This is
because the long distance physics of the Coulomb potential is dominant
in dimensions $\nu<3$, and the Lenard-Balescu equation is designed to
capture such long distance physics. I will write the Lenard-Balescu
equation in schematic form as
\begin{eqnarray}
  \frac{\partial f_a}{\partial t} + 
  {\bf v}_a \! \cdot \!{\bm\nabla}_x f_a
  &=& 
  {\sum}_b L_{ab}[f] ~~~~ : \nu < 3 \ ,
\label{LBEsimpnu}
\end{eqnarray}
where the kernel of the $\nu$-dimensional Lenard-Balescu equation
is the obvious generalization from three dimensions, 
\begin{eqnarray}
\nonumber
  L_{ab}[f]
  &=&
  \int\! \frac{d^\nu p_a}{(2\pi\hbar)^\nu}\, 
  \frac{d^\nu p_b}{(2\pi\hbar)^\nu}\,\frac{d^\nu k}{(2\pi)^\nu}
  \, 
  {\bm\nabla}_{\!\! p_a}
\!\cdot\!{\bf k}\,
  \bigg\vert \frac{e_a\, e_b}{k^2\, \epsilon({\bf k},{\bf v}_a\cdot{\bf k})}
  \bigg\vert^2 
  \pi\,\delta({\bf v}_a\!\cdot\!{\bf k}-{\bf v}_b\!\cdot\!{\bf k})
\\[5pt] 
  && \hskip6cm 
  \bigg\{
  {\bf k}\!\cdot\! {\bm\nabla}_{\!\!p_b}  \, 
  -
  {\bf k}\!\cdot\! {\bm\nabla}_{\!\!p_a}
  \bigg\}  f_a({\bf p}_a) \,  f_b({\bf p}_b) \ ,
\label{dedteigtAA}
\end{eqnarray}
where ${\bm\nabla}_{\!\! p_c}$ is the \hbox{$\nu$-dimensional} momentum
gradient. Reference~\cite{lifs} shows that the dielectric function of
a weakly to moderately coupled plasma is given by
\begin{eqnarray}
  \epsilon({\bf k},\omega) 
  = 
  1 + {\sum}_c \, \frac{e_c^2}{k^2} 
  \int \frac{d^\nu{\bf p}_c}{(2\pi\hbar)^\nu}\, 
  \frac{1}{\omega - {\bf k} \!\cdot\! 
  {\bf v}_c + i \eta}\, {\bf k} \cdot 
  {\bm\nabla}_{\!\! p_c}\, f_c({\bf p}_c) \,,
\label{epsilon}
\end{eqnarray}
where the prescription $ \eta \to 0^+ $ is implicit and defines the
correct retarded response. We can use (\ref{epsilon}) in
(\ref{dedteigtAA}) to the order in $g$ to which we are working.  The
sum in (\ref{epsilon}) is over all plasma components, and the velocity
${\bf v}_c={\bf p}_c/m_c$ appearing in the denominator is really an an
integration variable.  Therefore, when $\nu<3$, the rate
(\ref{dedtbenuab}) allows us to express
\begin{eqnarray}
  \frac{d{\cal E}_{ab}^\smLT}{dt}
  =
  \int \!\frac{d^{\,\nu}p}{(2\pi\hbar)^\nu}~\frac{p^2}{2 m_a}~
  L_{ab}[f] ~~~~ : ~~ \nu<3 \ .
\label{dedtltthree}
\end{eqnarray}
I have used a ``less than'' superscript to remind us that we should
calculate (\ref{dedtltthree}) in dimensions less than three.  

It is convenient to express the dielectric function in
terms of a complex function $F(v)$ defined by the relation
\begin{equation}
  k^2 \, \epsilon({\bf k} , {\bf k}\!\cdot\!{\bf v}) =
  k^2 + F(v \cos\theta) \ ,
\label{epsF}
\end{equation}
where $\theta$ is the angle between ${\bf k}$ and ${\bf v}$. The
engineering unit of the argument of $F$ is velocity, while the unit of
$F$ itself is wave-number squared. Expressions (\ref{epsilon}) and
(\ref{epsF}) imply the dispersion relation
\begin{equation}
  F(u)  =  \int_{-\infty}^{+\infty} \!\! dv \,
  \frac{\rho_{\rm total}(v)}{v - u - i \eta} \,,
\label{Fdisp}
\end{equation}
where the limit $\eta \to 0^+$ is understood, with the spectral 
weight being defined by
\begin{eqnarray}
\label{rhototdef}
  \rho_\text{total}(v)  
  &=& 
  {\sum}_c \, \rho_c(v)
\\[10pt]
\label{rhocdef}
  \rho_c(v) 
  &=&
  \kappa^2_c \,\left(\frac{\beta_c m_c}{2\pi}\right)^{1/2}
  v\, \exp\!\left\{-\frac{1}{2} \beta_c m_c v^2 \right\} \ .
\end{eqnarray}
For future use, we shall require the convenient relations 
\begin{eqnarray}
  F(-v) &=& F^*(v)
\label{fone}
\\[5pt]
  {\rm Im}\,F(v)
  &=&
  \frac{1}{2i}\Big[F(v)-F^*(v)\Big] = \pi\,\rho_\text{total}(v) \ .
\label{ftwo}
\end{eqnarray}
While we shall not require the real part of $F$, nor is there space
to compute this function, for completeness I shall record it here:
\begin{eqnarray}
  {\rm Re}\,F(v)
  &=& 
  \sum_b \kappa_b^2 
  \left[1 - 2 \sqrt{\frac{\beta_b m_b}{2}}\, v~
  {\rm daw}\left(\sqrt{\frac{\beta_b m_b}{2}}\,v 
  \right) \right] \ ,
\label{frq}
\end{eqnarray}
where the Dawson integral is defined by
\begin{eqnarray}
  {\rm daw}(x) \equiv \int_0^x dy\, 
  e^{y^2 - x^2} \ .
\end{eqnarray}

\subsection{\label{sec:rate} Calculating the Rate}

Returning to three dimensions for a moment, and dropping the species
index on (\ref{Ea}) for simplicity, the rate of change in the kinetic
energy density is simply given by
\begin{eqnarray}
  \frac{d {\cal E}}{d t} 
  =
  \int\! \frac{d^3 p}{(2\pi\hbar)^3}\,
  \frac{p^2}{2 m}\,
  \frac{\partial f}{\partial t} ({\bf p},t) 
\label{ratea}
\end{eqnarray}
The problem with a straightforward evaluation of (\ref{ratea}) in
three dimensions is that any potentially tractable kinetic equation
gives a logarithmically divergent result for the rate, either at short
or long distances, depending on the deficiencies of the particular
kinetic equation in hand. In principle, however, calculating the rate
is a well defined procedure: it is the approximation scheme employed
in finding the requisite distribution function that introduces a
divergence. In other words, if one {\em knew} the exact
single-particle distribution function $f({\bf p},t)$, then the rate
(\ref{ratea}) would be finite. However, the one-point function
\hbox{$f_1 = f({\bf p},t)$} of the BBGKY hierarchy can be known
exactly only by solving the entire set of coupled multi-particle
correlation functions exactly, an impossible feat. Hence, we must
approximate the exact distribution function by one obtained through
truncating the BBGKY kinetic equations.  Even worse, the truncation
process is rather subjective in that it depends upon the type of
physics one deems important; for example, truncation to the Boltzmann
equation is only useful if we can neglect long distance correlations
(which, in this problem, we cannot\footnote{
\footnoteskip
  In reducing BBGKY to the Boltzmann equation, we make the
  approximation that two-body collisions are uncorrelated, thereby
  allowing the replacement $f_2 \to f_1 \cdot f_1$ in the scattering
  kernel. However, the correlations described by $f_2$ and higher act
  back on $f_1$ to render the integrals in (\ref{ratea}) finite at
  long distances (and such correlations are neglected by the Boltzmann
  equation). }).
Conversely, truncation to the Lenard-Balescu equation captures the
long distance physics, but misses the short distance physics.  In a
nutshell, then, our problem is the following: to calculate the rate,
we require the exact distribution function of the full hierarchy of
kinetic equations, a problem we cannot hope to solve without a Quantum
Computer or a Mentat; we must therefore truncate the kinetic
equations, but this does violence to either the short or long distance
physics, thereby introducing spurious divergences.

A hint out of this Catch 22 comes from the following observation. The
truncation problem only occurs for the {\em Coulomb potential}, and
only then in {\em three} spatial dimensions.  As we saw in
(\ref{Vnu}), the Coulomb potential in $\nu$ dimensions is $V_\nu(r) =
C_\nu\, e_a e_b/r^{\nu-2}$, where $C_\nu=\Gamma(\nu/2-1)/4 \pi^{
\nu/2}$ is a geometric constant, and this form of the potential
renders the scattering kernels of the Boltzmann and Lenard-Balescu
equations finite, except for the single case $\nu=3$ (ironically, the
case of interest).  Let us therefore start with the exact BBGKY
hierarchy in $\nu$ spatial dimensions, with the understanding that
taking $\nu \ne 3$ is a regulating procedure, and that we must
eventually return to three dimensions. This procedure, however, is
robust enough to capture the correct physics as the limit $\nu \to 3$
is taken.

\subsubsection{\label{sec:reduction} Reduction of BBGKY}

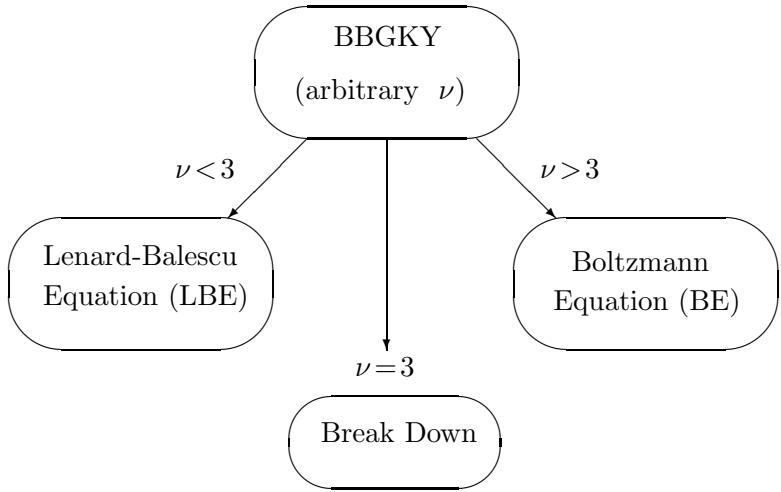
\begin{figure}[t]
\begin{picture}(120,120)(-75,0)
\put(-20,80){\oval(100,50)}
\put(-40,90){\text{BBGKY}}
\put(-55,70){(\text{arbitrary } $\nu$) }

\put(-50,55){\vector(-1,-1){30}}
\put(14,55){\vector(1,-1){30}}
\put(-100,40){$\nu\!<\!3$}\
\put(38,40){$\nu\!>\!3$}

\put(50,5){\text{Boltzmann}}
\put(43,-10){\text{Equation (BE)}}
\put(78,0){\oval(100,50)}

\put(-150,7){\text{Lenard-Balescu}}
\put(-150,-8){\text{Equation (LBE)}}
\put(-113,0){\oval(100,50)}

\put(-20,55){\vector(0,-1){80}}
\put(-32,-35){$\nu\!=\!3$}
\put(-45,-60){\text{Break Down}}
\put(-17,-60){\oval(80,35)}

\end{picture}
\vskip3cm 
\caption{\baselineskip12pt plus 1pt minus 1pt For $\nu\!>\!3$ the
``textbook derivation'' of the Boltzmann equation for a Coulomb
potential is rigorous; furthermore, the BBGKY hierarchy reduces to the
Boltzmann equation to leading order in $g$. A similar reduction from
the BBGKY hierarchy holds for the Lenard-Balescu equation in $\nu
\!<\! 3$, and the ``textbook derivation'' is also rigorous in these
dimensions. In $\nu=3$, the derivations of the Boltzmann and
Lenard-Balescu equations break down for the Coulomb potential.  }
\label{fig:bbgky}
\end{figure}

In the exact same manner as in the last few section, one can
generalize the BBGKY hierarchy to an arbitrary number of
dimensions. Furthermore, as we discussed in Lecture~I~\cite{bps1},
when the number of spatial dimension is greater than three, BBGKY
reduces to the Boltzmann equation (\ref{BEsimpnu}) and
(\ref{BEfeSigma}) to leading order in the plasma coupling $g$.
Conversely, when $\nu<3$ the BBGKY hierarchy reduces to the
Lenard-Balescu equation (\ref{LBEsimpnu}) and (\ref{dedteigtAA}) to
leading order in $g$. As discussed in the previous section, these
reduced kinetic equations (Boltzmann and Lenard-Balescu) are finite in
their respective dimensional regimes.  In other words, besides
rendering the truncation process finite, the physical utility of
keeping the dimension of space arbitrary is that in dimensions greater
than three, the leading order in $g$ behavior of BBGKY is just the
finite \hbox{$\nu$-dimensional} Boltzmann equation:
\begin{eqnarray}
  \text{BBGKY in}~\nu>3~\Rightarrow~
  \frac{\partial f_a}{\partial t} + 
  {\bf v}_a \! \cdot \!{\bm\nabla}_{\!\!x} f_a
  &=& 
  {\sum}_b B_{ab}[f]   
 ~~~\text{to LO in } g \ ,
\label{BEnu}
\end{eqnarray}
where $B_{ab}[f]$ is given by (\ref{BEfeSigma}). 
As discussed at length in Lecture~I, expression (\ref{BEnu}) is the
point at which the {\em physics} of dimensional continuation enters
the calculation: dimensions greater than three select for short
distance physics.  For $\nu>3$, the rate of energy transport from
plasma species $a$ to species $b$ is therefore given by the finite
expression (\ref{dedtgtthree}).  Turning now to dimensions less than
three, we have seen in Lecture~I that the leading order in $g$
behavior of BBGKY reduces to the Lenard-Balescu equation,
\begin{eqnarray}
  \text{BBGKY in}~\nu<3~\Rightarrow~
  \frac{\partial f_a}{\partial t} + 
  {\bf v}_a \! \cdot \!{\bm\nabla}_{\!\!x} f_a
  &=& 
  {\sum}_b L_{ab}[f]   
 ~~~\text{to LO in } g \ ,
\label{LBEnu}
\end{eqnarray}
where the finite $\nu$-dimensional scattering kernel $L_{ab}[f]$ is
given by (\ref{dedteigtAA}). Again, the physical content of
dimensional continuation enters at this stage: dimensions less than
three select for long distance physics.  This dimensional reduction of
BBGKY is illustrated in Fig.~\ref{fig:bbgky}.

From the rate equations (\ref{kequgen})--(\ref{dedtbenuab}), in
dimensions $\nu>3$ expression (\ref{BEnu}) gives the rate of energy
transport from species $a$ to species $b$ as
\begin{eqnarray}
  \frac{d {\cal E}_{ab}^\smGT}{d t} 
  =
  \int\! \frac{d^\nu p_a}{(2\pi\hbar)^\nu}\,
  \frac{p_a^2}{2 m_a}\, B_{ab}[f]   
 ~~\text{to LO in } g \ ,
\label{nulargeBE}
\end{eqnarray}
which, as we shall see, takes the form 
\begin{eqnarray}
  \frac{d{\cal E}_{ab}^\smGT}{dt}
  &=& 
  H(\nu)\,\frac{g^2}{\nu-3} 
  +
  {\cal O}(\nu-3) 
  ~~~  \text{to LO in $g$ when }\nu > 3 \ .
\label{dedtonecal}
\end{eqnarray}
We have omitted the species indices from $H(\nu)$ for simplicity, and
we shall calculate this quantity in Section~\ref{sec:BE}.  Rather than
a logarithmically divergent result, we obtain a finite answer
involving a simple pole of the form $1/(\nu-3)$, which of course is
the origin of the divergence in three dimensions. Since we will
eventually return to $\nu=3$, there is no need to calculate the ${\cal
O}(\nu-3)$ terms in (\ref{dedtonecal}), as these terms vanish when
$\nu \to 3$.  Similarly, from (\ref{LBEnu}) the corresponding rate in
energy transport from species $a$ to $b$ is therefore
\begin{eqnarray}
  \frac{d {\cal E}_{ab}^\smLT}{d t} 
  =
  \int\! \frac{d^\nu p_a}{(2\pi\hbar)^\nu}\,
  \frac{p_a^2}{2 m_a}\, L_{ab}[f]   
 ~~~\text{to LO in } g \ .
\label{nusmallLB}
\end{eqnarray} 
From the calculation in Sec.~\ref{sec:LBE}, we shall find
\begin{eqnarray}
  \frac{d{\cal E}_{ab}^\smLT}{dt}
  &=&
  G(\nu)\, \frac{g^{\nu-1}}{3-\nu} 
  + {\cal O}(3-\nu) 
  ~~~ \text{to LO in $g$ when } \nu < 3 \ ,
\label{dedttwocal}
\end{eqnarray}
where we have omitted the species indices from $G(\nu)$ to save
writing.  Note that the leading behavior of both (\ref{dedtonecal})
and (\ref{dedttwocal}) is formally of order $g^2$ in three dimensions,
illustrating that neither short nor long distance physics dominates in
$\nu=3$, but rather, that ultraviolet and infrared length scales
contribute to the same order in three dimensions. It is a property of
the Coulomb potential itself that $\nu=3$ is the fulcrum around which
the short and long distance physics pivot.  Using the appropriate
Coulomb potential (\ref{Vnu}) for $V_\nu(r)$ in the scattering kernels
of (\ref{nulargeBE}) and (\ref{nusmallLB}), the integrals now
converge, and they are calculated exactly in Sections~7 and 8 of
BPS~\cite{bps}. In these notes we shall calculate them in
Sections~\ref{sec:BE}~and~\ref{sec:LBE}.

\subsubsection{Obtaining Next-to-Leading Order from Leading Order}

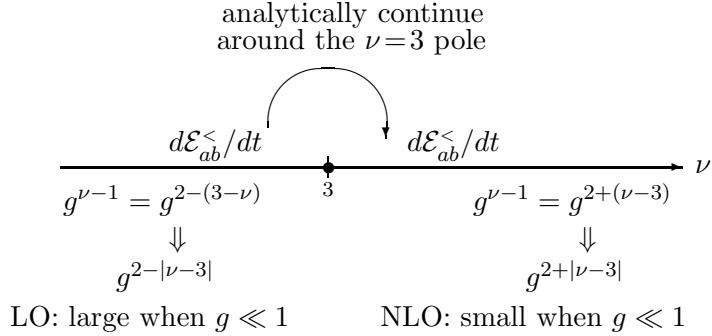
\begin{figure}[t]
\begin{picture}(120,120)(-75,0)
\put(-101,60){\vector(1,0){235}}
\put(139,58){$\nu$}

\put(0,56){\line(0,1){8}}
\put(-2.3,57.3){$\bullet$}
\put(-2.0,49){${\scriptstyle 3}$}

\put(-60,66){$d{\cal E}_{ab}^\smLT/dt$}
\put(-100,44){$g^{\nu-1}=g^{2-(3 - \nu)}$}
\put(-60,30){$\Downarrow$}
\put(-80,15){$g^{2-\vert \nu-3 \vert}$}
\put(-120,0){LO: large when $g \ll 1$}

\put(30,66){$d{\cal E}_{ab}^\smLT/dt$}
\put(55,44){$g^{\nu-1}=g^{2+(\nu-3)}$}
\put(95,30){$\Downarrow$}
\put(75,15){$g^{2+\vert \nu-3 \vert}$}
\put(20,0){NLO: small when $g \ll 1$}

\put(0,75){\oval(45,45)[t]}
\put(22,72){\vector(0,-1){1}}
\put(-40,115){analytically continue}
\put(-42,105){around the $\nu\!=\!3$ pole}

\end{picture}
\caption{\captionskip The analytic continuation of $d{\cal E }_{
ab}^\smLT/dt$ from $\nu<3$ to the region $\nu>3$ in the complex
$\nu$-plane. The same expression can be used for $d{\cal E }_{
ab}^\smLT/dt$ throughout the complex plane since the pole at
$\nu=3$ can easily be avoided. The quantity $d{\cal E }_{ab}^\smLT
/ dt \sim g^{2+(\nu-3)}$ is leading order in $g$ for $\nu <
3$. However, upon analytically continuing to $\nu>3$ we find that
$d{\cal E}_{ab}^\smLT/dt\sim g^{2+\vert \nu-3\vert}$ is
next-to-leading order in $g$ relative to $d{\cal E}_{ab}^\smGT/dt
\sim g^2$.  }
\label{fig:NLOdedt}
\end{figure}

In three dimensions, or in the limit $\nu \to 3$, we still are still
plagued by the long and short distance divergences from the simple
poles in (\ref{dedtonecal}) and (\ref{dedttwocal}), a problem we must
now confront if we are to obtain a meaningful result.  As described in
Lecture~I, to compare the rates (\ref{dedtonecal}) and
(\ref{dedttwocal}), we must analytically continue one or the other to
a {\em common} value of the dimension $\nu$.  Analytically continuing
the spatial dimension makes sense because we can view the quantities
$d {\cal E }^\smGT_{ab}/dt$ and $d {\cal E }^\smLT_{ab}/dt$ as
functions of a complex parameter $\nu$, even though they were only
calculated for positive integer values of $\nu$. This is analogous to
analytically continuing the factorial function on the positive
integers to the Gamma function on the complex plane.  For
definiteness, I will analytically continue (\ref{dedttwocal}) to
$\nu>3$, in which case the $g$-dependence becomes subleading relative
to the $g^2$ dependence of (\ref{dedtonecal}). The analytic
continuation of (\ref{dedttwocal}) takes the same functional form for
any $\nu \in \mathbb{C}$, but in this section I will write the
analytic continuation as
\begin{eqnarray}
  \frac{d{\cal E}_{ab}^\smLT}{dt}
  &=&
  -G(\nu)\, \frac{g^{2+(\nu-3)}}{\nu-3} 
  +
  {\cal O}(\nu-3) 
  ~~~\text{to NLO in $g$ when } \nu > 3 \ .
\label{NLOgterm}
\end{eqnarray}
Since we are now working in the regime $\nu-3>0$, I have written the
exponent of $g$ in a form to emphasize that $g^2 \gg g^{\nu-1}$ when
$g \ll 1$ and $\nu>3$. There are no terms with powers of $g$
intermediate $g^2$ and $g^{\nu-1}$, so the analytic continuation of
$d{\cal E}_{ab}^\smLT/dt$ to parameters $\nu>3$ is not only
subleading in $g$, but it is indeed next-to-leading order relative to
(\ref{dedtonecal}).  This is illustrated in Fig.~\ref{fig:NLOdedt}.

In the rates (\ref{dedtonecal}) and (\ref{dedttwocal}), we need to
work consistently only to linear order in the small parameter
$\epsilon = \nu-3$; therefore, we should expand $H(\nu)$ and $G(\nu)$
to first order in $\epsilon$, allowing us to write
\begin{eqnarray}
  H(\nu) 
  &=&
  -A + \epsilon \,H_1 + {\cal O}(\epsilon^2)
\label{Hexp}
\\[5pt]
  G(\nu) 
  &=&
  -A + \epsilon \,G_1 + {\cal O}(\epsilon^2) \ .
\label{Gexp} 
\end{eqnarray}
It is crucially important here that $H(\nu)$ and $G(\nu)$ give the
same value at $\nu=3$, a term that I have called $A$ in (\ref{Hexp})
and (\ref{Gexp}), otherwise the divergent poles will not cancel. In
Section~\ref{sec:calmain}, we shall calculate the coefficients
$H(\nu)$ and $G(\nu)$, and we will indeed explicitly see that
$H(\nu\!=\!3)$ and $G(\nu\!=\!3)$ are equal. We will also calculate
$H_1 \equiv H^\prime(\nu\!=\!3)$ and $G_1 \equiv G^\prime(\nu\!=\!3)$
in closed form, thereby providing an exact result for the rate to
leading and next-to-leading order in $g$.

Finally, upon writing $g^\epsilon = \exp\{\epsilon \ln g\}$ in
(\ref{NLOgterm}), and expanding to first order in $\epsilon$, we find
\begin{eqnarray}
  \frac{g^\epsilon}{\epsilon}
  = 
  \frac{1}{\epsilon} 
  +
  \ln g + {\cal O}(\epsilon) \ .
\label{gexp}
\end{eqnarray}
This is where the nonanalyticity in $g$ arises, {\em i.e.} the $\ln g$
term, and we can now express the rates (\ref{dedtonecal}) and
(\ref{NLOgterm}) as
\begin{eqnarray}
  \frac{d{\cal E}_{ab}^\smGT}{dt}
  &=& 
  -
  \frac{A}{\nu-3} \, g^2
  +
  H_1\,g^2
  +
  {\cal O}(\nu-3;g^3)
  \hskip2.7cm ~\nu>3 
\label{LOcalcA}
\\[5pt]
  \frac{d{\cal E}_{ab}^\smLT}{dt}
  &=& 
    \phantom{-}
  \frac{A}{\nu-3} \, g^2 
  - 
  G_1 \,g^2
  - 
  A\, g^2 \ln g 
  +
  {\cal O}(\nu-3;g^3) 
  \hskip0.75cm ~\nu>3 
  \ .
\label{LOcalcB}
\end{eqnarray}
These expressions hold in the common dimension $\nu > 3$, and
to obtain the leading and next-to-leading order result in three
dimensions, we add and take the limit:
\begin{eqnarray}
  \frac{d{\cal E}_{ab}}{dt}
  =
  \lim_{\nu \to 3^+}
  \left[
  \frac{d{\cal E}_{ab}^\smGT}{dt}
  +
  \frac{d{\cal E}_{ab}^\smLT}{dt}
  \right] + {\cal O}(g^3) 
  =
  - A g^2 \ln g + B g^2 + {\cal O}(g^3) \ ,
\label{dedtorderg3}
\end{eqnarray}
with $B=H_1-G_1$. Compare this with the rate (\ref{dedtNLO}), or the
alternative expression (\ref{lngsqu}). In this way, BPS calculated the
energy exchange rate from Coulomb interactions between plasma species,
accurate to leading order and next-to-leading order in $g$.

\pagebreak
\section{\label{sec:calmain} Calculating the Rate to Subleading Order}

We now perform the real calculation of the energy exchange rate
between electrons and ions. BPS~\cite{bps} considered the general
case, finding ${\cal C}_{ab}$ in (\ref{Cabdef}) for any collection of
plasma species in any quantum regime. In these notes, however, I will
only consider the extreme quantum limit valid at high temperatures.
This is the case of most general interest, and it is also the case in
which the algebra simplifies considerably. We shall also take the
electrons to be in equilibrium with themselves at temperature $T_e$
and the ions in equilibrium with themselves at temperature $T_\smI$,
another situation of general interest. Upon summing over the ions, the
rate equation then becomes
\begin{eqnarray}
  \frac{d {\cal E}_{e\smI}}{dt} 
  = 
  - {\cal C}_{e\smI}\,(T_e - T_\smI) \ ,
\label{dEdtCeIdef}
\end{eqnarray}
where the collective rate coefficient that we shall calculate is
\begin{eqnarray}
{\cal C}_{e\smI} = {\sum}_i {\cal C}_{ei} \ .
\label{CeIdef}
\end{eqnarray}

\subsection{\label{sec:BE} Boltzmann-Equation: Short-Distance Physics}

We now work in $\nu>3$ dimensions where the short-distance physics
dominates.  To calculate the rate of change of the electron
distribution, we will employ the Boltzmann equation with two-body
quantum effects in the Born approximation.  The rate of energy
exchange from the electrons to the ions is therefore,
\begin{eqnarray}
  \frac{\partial{\cal E}_{e\smI}^\smGT}{\partial t}
  =
  \int \frac{d^\nu p_e}{(2\pi\hbar)^\nu}\, 
  \frac{p_e^2}{2 m_e}\,
  \frac{\partial f_e ({\bf p}_e)}{\partial t} 
  =
  \sum_i\int \frac{d^\nu p_e}{(2\pi\hbar)^\nu}\, 
  \frac{p_e^2}{2 m_e}\,
  B_{e i}[f] \ ,
\label{dedteiA}
\end{eqnarray}
where we have taken $a=e$ and $b=i$ in (\ref{BEsimpnu}) and
(\ref{BEfe}), and the electron and ion distribution functions are
given by (\ref{defFiA}). Using the crossing symmetries ${\bf p}_e
\leftrightarrow {\bf p}_e^\prime$ and ${\bf p}_i \leftrightarrow {\bf
p}_i^\prime$ of the scattering amplitude $T_\smB$ in (\ref{BEfe}), the
rate of energy exchange from the electrons to an ion species $i$ can
be written
\begin{eqnarray}
\nonumber
  \frac{\partial{\cal E}_{e i}^\smGT}{\partial t}
  &=&
  \int 
  \frac{d^\nu p_e^\prime}{(2\pi\hbar)^\nu}\,
  \frac{d^\nu p_i^\prime}{(2\pi\hbar)^\nu}\,
  \frac{d^\nu p_e}{(2\pi\hbar)^\nu}\, 
  \frac{d^\nu p_i}{(2\pi\hbar)^\nu}\,\big\vert T_\smB\big\vert^2 ~
  \frac{p_e^{\prime\,2}-p_e^2}{2 m_e}\, f_e({\bf p}_e) f_i({\bf p}_i) 
\\ &&
  (2\pi\hbar)^\nu\,\delta^\nu\!\Big( {\bf p}_i^\prime + 
  {\bf p}_e^\prime  - {\bf p}_i - {\bf p}_e \Big)\,
  (2\pi\hbar)\, \delta\!\left( 
  \frac{p_e^{\prime\,2}-p_e^2}{2m_e} 
  + 
  \frac{p_i^{\prime\,2}-p_i^2}{2m_i} \right) \ ,
\label{dedteiBB}
\end{eqnarray}
where we dropped the gradient term in (\ref{BEfe}) because of spatial
uniformity. Summing over all ions in (\ref{dedteiBB}) gives the total
rate (\ref{dedteiA}). We define the momentum transfer ${\bf q}$ and
the average momentum $\bar{\bf p}$ of the initial and final
electron momentum,
\begin{eqnarray}
  {\bf q} 
  &\equiv& 
  {\bf p}_e^\prime - {\bf p}_e 
  =
  {\bf p}_i - {\bf p}_i^\prime
\\[5pt]
  \bar{\bf p} 
  &\equiv& 
  \frac{1}{2}\,\Big[\,{\bf p}_e^\prime + {\bf p}_e \Big] \ .
\end{eqnarray}
Upon performing the $p_i^\prime$-integral using the momentum
conserving delta-function to set 
\begin{eqnarray}
  {\bf p}_i^\prime 
  =
  {\bf p}_i + {\bf p}_e - {\bf p}_e^\prime
  =
  {\bf p}_i - {\bf q} \ ,
\end{eqnarray}
and expressing the electron momenta in terms of
${\bf q}$ and $\bar{\bf p}$, 
\begin{eqnarray}
  {\bf p}_e^\prime 
  &=& 
  \bar{\bf p} + \frac{1}{2}\,{\bf q}
  ~~~~~~
  {\bf p}_e 
  =
  \bar{\bf p} - \frac{1}{2}\,{\bf q}\ ,
\end{eqnarray}
we can simplify (\ref{dedteiBB}) to read 
\begin{eqnarray}
\nonumber
  \frac{\partial{\cal E}_{ei}^\smGT}{\partial t}
  &\!=\!& \!
  \int \!\!
  \frac{d^\nu \bar p}{(2\pi\hbar)^\nu}\,
  \frac{d^\nu q}{(2\pi\hbar)^\nu}\, 
  \big\vert T_\smB(q)\big\vert^2\, 
  \frac{m_i}{m_e}\,\bar{\bf p}\!\cdot\! {\bf q}~
  f_e(\bar{\bf p} -{\bf q}/2) \,\smTimes
\\[5pt] && 
  \int \frac{d^\nu p_i}{(2\pi\hbar)^\nu}\,
  (2\pi\hbar)\delta\!\left({\bf p}_i\!\cdot\! {\bf q} -
  \frac{{m_i}}{2m_e}\,\bar{\bf p}\!\cdot\!{\bf q}-\frac{1}{2}\,
  q^2 \right)\,f_i({\bf p}_i) \ .
\label{dedteiDD}
\end{eqnarray}
We have used the fact that the energy conserving delta function and
the energy loss factor become,
\begin{eqnarray}
  \frac{p_e^{\prime\,2}-p_e^2}{2 m_e}
  &=&
  \frac{1}{m_e}\,\bar{\bf p}\!\cdot\! {\bf q} 
\\[5pt]
  \delta\!\left(\frac{p_e^{\prime\,2}-p_e^2}{2m_e} + 
  \frac{p_i^{\prime\,2}-p_i^2}{2m_i} \right)
  &=&
  \delta\!\left(\frac{\bar{\bf p}\!\cdot\!{\bf q}}{m_e}- 
  \frac{{\bf p}_i\!\cdot\! {\bf q}}{2 m_i} +\frac{q^2}{2m_i}  
  \right) \ .
\end{eqnarray}

We now perform the $p_i$-integration. Since we will find a similar
integral in the next section, I will perform a more general
calculation here. There will be times when we need to integrate a
Gaussian and a delta-function, which I will write
as 
\begin{eqnarray}
  \int\frac{d^\nu p_b}{(2\pi\hbar)^\nu}\, 
  \delta(\hat{\bf k}\!\cdot\!{\bf v}_b-V)\, 
  e^{-\beta_b E_b}
  &=&
  \hbar\, \beta_b \left(\frac{1}{\lambda_b}\right)^{\nu+1} \!\!
  e^{-\frac{1}{2}\,\beta_b m_b\, V^2} 
  =
  \frac{m_b}{2\pi\hbar}\, \lambda^{1-\nu}\,
  e^{-\frac{1}{2}\,\beta_b m_b\, V^2}  ,
\label{intdelfiexp}
\end{eqnarray}
where $\hat{\bf k}$ is a fixed unit vector (typically another
integration variable), and $V$ is a scalar independent of $p_b=
m_b{\bf v}_b$. The integral (\ref{intdelfiexp}) will be required in
several places throughout the text, so we will perform the calculation
here.  Since $\mathbb{V}=\hat{\bf v}_b\cdot\hat{\bf k}$ defines the
component of ${\bf v}_b$ parallel to $\hat{\bf k}$, will decompose the
integration variables ${\bf v}_b$ into parallel and normal components
\begin{eqnarray}
  {\bf v}_b
  = 
  {\bf v}_\smPerp + ({\bf v}_b\!\cdot\! \hat{\bf k})\,\hat{\bf k} 
  = 
  {\bf v}_\smPerp + \mathbb{V}\,\hat{\bf k} \ .
\end{eqnarray}
Since $v_b^2 = v_\smPerp^2 + \mathbb{V}^2$, we can write 
\begin{eqnarray}
\nonumber
  &&
  \int\!\!\frac{d^\nu p_b}{(2\pi\hbar)^\nu}\, 
  \delta({\bf v}_b\!\cdot\!\hat{\bf k}-V)\, 
  e^{-\ell \beta_b\, p_\ell^2/2 m_b}
\\[5pt]
  && =
  \left(\frac{m_b}{2\pi\hbar}\right)^\nu \!\!
  \int d^{\nu-1} v_\smPerp 
  e^{-\ell\frac{1}{2} \beta_b m_b {v_\smPerp}^2}
  \cdot 
  \int_{-\infty}^\infty d \mathbb{V} \,\delta(\mathbb{V}-V)\, 
  e^{-\ell\frac{1}{2} \beta_b m_b \mathbb{V}^2 }
\\[5pt]
  && =
  \left(\frac{m_b}{2\pi\hbar}\right)^\nu 
  \left(\frac{2\pi}{\ell \beta_b m_b}\right)^{(\nu-1)/2}
  \cdot 
  e^{-\ell\frac{1}{2} \beta_b m_b V^2 } \ ,
\end{eqnarray}
and this yields expression (\ref{intdelfiexp}). As an application,
we will frequently run across an integral 
\begin{eqnarray}
  \int\frac{d^\nu p_b}{(2\pi\hbar)^\nu}\, 
  \delta({\bf v}_b\!\cdot\!\hat{\bf k}-V)\, f_b(p_b) 
  =
  n_b \left( \frac{\beta_b m_b}{2\pi}\right)^{1/2} 
  e^{-\frac{1}{2}\,\beta_b m_b\, V^2} \ ,
\label{intdelfi}
\end{eqnarray}
which follows directly from (\ref{intdelfiexp}). Using the integral
(\ref{intdelfi}), and taking (\ref{defFiA}) for $f_e(\bar{\bf p}- {\bf
q}/2)$ gives
\begin{eqnarray}
\nonumber
  \frac{\partial{\cal E}_{ei}^\smGT}{\partial t}
  &=&
  \frac{n_e n_i}{m_e}\,
  \left(2\pi \beta_i m_i \right)^{1/2} \lambda_e^\nu
  \int \!
  \frac{d^\nu \bar p}{(2\pi\hbar)^\nu}\,
  \frac{d^\nu q}{(2\pi\hbar)^\nu}\, 
  \hbar\, \big\vert T_\smB(q)\big\vert^2\,
  \bar{\bf p}\!\cdot\! \hat{\bf q}\,
\\[5pt] && \hskip-2cm 
  \exp\!\bigg\{\!
  -\frac{\beta_i m_i}{2 m_e^2}\,\bigg[
  \left(\bar{\bf p}\!\cdot\!\hat{\bf q}
  \right)^2 
  +
  \frac{m_e}{m_i}
  \left(1 - \frac{\beta_e}{\beta_i}\right)q\,
  \bar{\bf p}\!\cdot\!\hat{\bf q}
  \bigg]
  -
  \frac{\beta_e}{2m_e}\bigg[\,
  {\bar p}^{\,2} 
  + 
  \left(1+\frac{m_e \,\beta_i}{m_i\,\beta_e}\,\right)\!\frac{q^2}{4}
  \bigg]
  \bigg\} \,.
\label{dedteiJJ}
\end{eqnarray}
When $\beta_e=\beta_i$, the linear term in the exponential involving
$\bar{\bf p} \cdot {\bf q}$ vanishes.  The integrand is even in both
$\bar{\bf p}$ and $\bar{\bf q}$, except for the prefactor $\bar{\bf
p}\cdot\hat{\bf q}$; therefore, keeping ${\bf q}$ fixed and
integrating over $\bar{\bf p}$ gives zero,
\begin{eqnarray}
\nonumber
  \frac{\partial{\cal E}_{ei}^\smGT}{\partial t}
  \Bigg\vert_{\beta_e=\beta_i} \!\! = 0 \ ,
\end{eqnarray}
as it should for equal electron and ion temperatures.

We now examine the general case when the electron and ion temperatures
differ. Completing the square for the terms in the first
square-brackets suggests changing variables to 
\begin{eqnarray}
  {\bar{\bf p}}^{\prime}
  =
  \bar{\bf p}
  +
  \frac{m_e}{m_i}
  \left(1 - \frac{\beta_e}{\beta_i}\right)
  \frac{{\bf q}}{2} \ ,
\label{pbarprimeA}
\end{eqnarray}
and dropping $q^2$-terms that are down by relative factors of
$m_e/m_i$ gives 
\begin{eqnarray}
\nonumber
  \frac{\partial{\cal E}_{ei}^\smGT}{\partial t}
  &=&
  \frac{n_e\, n_i}{m_e}\,
  \left(2\pi \beta_i m_i \right)^{1/2} \lambda_e^\nu
  \int \!
  \frac{d^\nu {\bar p}^{\,\prime}}{(2\pi\hbar)^\nu}\,
  \frac{d^\nu q}{(2\pi\hbar)^\nu}\, 
  \hbar\, \big\vert T_\smB(q)\big\vert^2\,
  \bigg[  {\bar{\bf p}}^{\,\prime} \!\cdot\! \hat{\bf q}
  + \frac{m_e}{m_i}
  \left(\frac{\beta_e}{\beta_i}-1\right)\!\frac{q}{2} 
  \bigg]
\\[5pt] && \hskip2cm 
  \exp\!\bigg\{\!
  -\frac{\beta_i m_i}{2 m_e^2}\, 
  \left({\bar{\bf p}}^{\,\prime}\!\cdot\!\hat{\bf q}\right)^2 
  -
  \frac{\beta_e}{2m_e}\bigg[\,
  {\bar p}^{\,\prime\,2}_\smPerp
  +   
  \left({\bar{\bf p}}^{\,\prime}\!\cdot\!\hat{\bf q}\right)^2 
  +
  \frac{q^2}{4}
  \bigg]
  \bigg\} \ ,
\label{dedteiKK}
\end{eqnarray}
where we have expanded the new integration variable ${\bar{\bf
p}}^\prime$ into normal and parallel components
\begin{eqnarray}
  {\bar{\bf p}}^\prime
  =
  {\bar{\bf p}}_\smPerp^\prime 
  + 
  \bar{\mathbb{P}}^\prime\, \hat{\bf q} 
  ~~~\text{with}~~ 
  \bar{\mathbb{P}}^\prime\,
  =
  {\bar{\bf p}}^\prime\!\cdot\!\hat{\bf q} \ .
\label{pbarprimedecA}
\end{eqnarray}
Finally, we note that the term $\left({\bar{\bf p}}^{\,\prime}
\!\cdot\! \hat{\bf q}\right)^2$ in the square brackets of the
exponential is down by a factor $m_e/m_i$ relative to the first 
such term, and we can write (\ref{dedteiKK}) as
\begin{eqnarray}
\nonumber
  \frac{\partial{\cal E}_{ei}^\smGT}{\partial t}
  &=&
  \frac{n_e\, n_i}{m_e}\,
  \left(2\pi \beta_i m_i \right)^{1/2} \lambda_e^\nu
  \int\frac{d \mathbb{P}}{2\pi\hbar}\, 
  \int 
  \frac{d^{\nu-1} {\bar p^{\,\prime}}_\smPerp}{(2\pi\hbar)^{\nu-1}}\,
  \frac{d^\nu q}{(2\pi\hbar)^\nu}\, 
  \hbar\, \big\vert T_\smB(q)\big\vert^2\,
\\[5pt] && 
  \bigg[ \mathbb{P}^\prime
  +\frac{m_e}{m_i}
  \left(\frac{\beta_e}{\beta_i}-1\right)\!\frac{q}{2} 
  \bigg]
  \exp\!\bigg\{\!
  -\frac{\beta_i m_i}{2 m_e^2}\, \mathbb{P}^{\prime\,2}
  -
  \frac{\beta_e}{2m_e}\bigg[\,
  {\bar p}^{\,\prime\,2}_\smPerp
  +   
  \frac{q^2}{4}
  \bigg]
  \bigg\} \ .
\label{dedteiMM}
\end{eqnarray}
The linear $\mathbb{P}^\prime$ term in the prefactor integrates to
zero, and we arrive at
\begin{eqnarray}
\nonumber
  \frac{\partial{\cal E}_{ei}^\smGT}{\partial t}
  &=&
  \frac{n_e\, n_i}{2 m_i}\,
  \left(2\pi \beta_i m_i \right)^{1/2} \lambda_e^\nu
  \left(\frac{\beta_e}{\beta_i}-1\right)
  \int\frac{d \mathbb{P}}{2\pi\hbar}\, 
  \int 
  \frac{d^{\nu-1} {\bar p^{\,\prime}}_\smPerp}{(2\pi\hbar)^{\nu-1}}\,
  \frac{d^\nu q}{(2\pi\hbar)^\nu}\, 
  \hbar\, \big\vert T_\smB(q)\big\vert^2\,
\\[5pt] && \hskip5cm 
  q\,
  \exp\!\bigg\{\!
  -\frac{\beta_i m_i}{2 m_e^2}\, \mathbb{P}^{\prime\,2}
  -
  \frac{\beta_e}{2m_e}\bigg[\,
  {\bar p}^{\,\prime\,2}_\smPerp
  +   
  \frac{q^2}{4}
  \bigg]
  \bigg\} \ .
\label{dedteiNN}
\end{eqnarray}
The integrand is now Gaussian in the variables $\mathbb{P}^\prime$ and
${\bar {\bf p}}^{\,\prime\,2}_\smPerp$, and we find 
\begin{eqnarray}
  (2\pi\beta_i m_i)^{1/2}\int_{-\infty}^\infty
  \frac{d {\bar{\mathbb{P}}^{\,\prime}}}{2\pi\hbar}
  \exp\!\bigg\{\!-\frac{\beta_i m_i}{2 m_e^2}\,
  \bar{\mathbb{P}}^{\prime\,2}
  \bigg\}
  &=& 
  \frac{m_e}{\hbar} 
\\[10pt] 
  \lambda_e^\nu
  \int \frac{d^{\nu-1} {\bar
  p^{\,\prime}}_\smPerp}{(2\pi\hbar)^{\nu-1}}\, \exp\!\bigg\{
  \frac{\beta_e}{2 m_e} \,{\bar p}^{\,\prime\,2}_\smPerp \bigg\} 
  &=&
  \lambda_e \ .
\end{eqnarray}
Substituting back into (\ref{dedteiNN}) gives
\begin{eqnarray}
  \frac{\partial{\cal E}_{ei}^\smGT}{\partial t}
  &=&
  \frac{n_e\, n_i}{2 m_i}\,
  \frac{\lambda_e\,m_e}{\hbar} \,
  \underbrace{~~
  \left(\frac{\beta_e}{\beta_i}-1\right)
  ~~}_{\beta_e\big(T_i - T_e\big)}\,
  \int 
  \frac{d^\nu q}{(2\pi\hbar)^\nu}\, 
  \hbar\, \big\vert T_\smB(q)\big\vert^2\,
  q\,
  \exp\!\bigg\{\!
  -
  \frac{\beta_e}{2m_e}\,\frac{q^2}{4}
  \bigg\} \ .
\label{dedteiOO}
\end{eqnarray}
Up to this point, the two-body scattering amplitude $T$ could have 
been general, but we now explicitly employ the Born approximation
\begin{eqnarray}
  T_\smB = \hbar\,\frac{e e_i}{q^2} \ ,
\end{eqnarray}
so that 
\begin{eqnarray}
  \frac{\partial{\cal E}_{ei}^\smGT}{\partial t}
  =
  \kappa_e^2
  \Big(\frac{1}{2}\,\lambda_e\,m_e\,  \hbar^2 \Big)
  \omega_i^2
  \Big(T_i - T_e \Big)
  \int 
  \frac{d^\nu q}{(2\pi\hbar)^\nu}\, 
  \frac{1}{q^3}\,
  \exp\!\bigg\{\!
  -
  \frac{\beta_e}{2m_e}\,\frac{q^2}{4}
  \bigg\} 
  \ ,
\label{dedteiPP}
\end{eqnarray}
where the ion plasma frequency is $\omega_i^2= e_i^2\, n_i/m_i$ and
the electron Debye wave-number is $\kappa_e^2 = e^2 n_e/T_e$. 

Another expression we will encounter is the gamma-function,
\begin{eqnarray}
  \Gamma(z) = \int_0^\infty \!du\, u^{z-1}\,e^{-u}
  ~~~{\rm Re}(z)>0 \ ,
\label{GammaDef}
\end{eqnarray}
The gamma-function (\ref{GammaDef}) provides a nice trick
for calculating integrals of the form
\begin{eqnarray}
  a^{-n} \!=\! \frac{1}{\Gamma(n)}
  \int_0^\infty \! ds\, s^{n-1}\, e^{-a\, s} 
\label{gammatrickA} \ .
\end{eqnarray}
In doing calculations, we will often need to take the norm of a 
vector to some power, such as $\vert {\bf q}\vert^{-(\nu-3)/2}$. 
For example, we can use (\ref{gammatrickA}) to exponentiate the 
norm into a more easily handled Gaussian by writing $\vert {\bf 
q}\vert = ({\bf q}\cdot {\bf q})^{1/2}$, so that
\begin{eqnarray}
  \vert {\bf q}\vert^{-m} \!=\! \frac{1}{\Gamma(m/2)}
  \int_0^\infty \! ds\, s^{m/2-1}\, e^{-s\, {\bf q}\cdot{\bf q}} 
\label{gammatrickB} \ .
\end{eqnarray}
We can now perform the $q$-integral to give
\begin{eqnarray}
  \int \!
  \frac{d^\nu q}{(2\pi\hbar)^\nu}\, q^{-3}\,
  e^{-\beta_e\,q^2/8m_e}
  &=&
  \frac{\Omega_{\nu-1}}{(2\pi\hbar)^\nu}\int_0^\infty \!\! 
  dq \, q^{\nu-4} e^{-\beta_e\,q^2/8m_e}
  ~~~~~ : s = \frac{\beta_e}{8 m_e}\, q^2
\\[5pt]
  &=&
  \frac{1}{(2\pi\hbar)^\nu}\,
  \frac{2\pi^{\nu/2}}{\Gamma(\nu/2)}
  \left(\frac{8m_e}{\beta_e}\right)^{(\nu-3)/2}
  \frac{1}{2}
  \int_0^\infty \!\! ds \, s^{(\nu-5)/2}\, e^{-s}
\\[5pt]
  &=&
  \frac{1}{4\pi^2\,\hbar^3}\,
  \frac{\Gamma(3/2)}{\Gamma(\nu/2)}\,
  \left(\frac{4}{\lambda_e^2}\right)^{(\nu-3)/2}
  \Gamma\left( \frac{\nu-3}{2}\right) \ ,
\label{intqthree}
\end{eqnarray}
where $\Gamma(3/2)=\sqrt{\pi}/2$, and substituting  this back
into (\ref{dedteiPP}) allows us to express
\begin{eqnarray}
  \frac{\partial{\cal E}_{ei}^\smGT}{\partial t}
  &=&
  \frac{\kappa_e^2}{2\pi}
  \Big(
  \frac{\lambda_e\,m_e}{2\pi \hbar} \Big) \omega_i^2
  \Big(T_i - T_e \Big)
  \frac{\Gamma(3/2)}{\Gamma(\nu/2)}\,
  \frac{1}{2}\,
  \left(\frac{4}{\lambda_e^2}\right)^{(\nu-3)/2}
  \Gamma\left( \frac{\nu-3}{2}\right) \ .
\label{dedteiQQ}
\end{eqnarray}
Note that 
\begin{eqnarray}
  \frac{\lambda_e\,m_e}{2\pi\hbar} 
  =
  \left(\frac{\beta_e m_e}{2\pi}\right)^{1/2} \ ,
\end{eqnarray}
and therefore the rate coefficient becomes, upon dividing by
the temperature difference and then summing over the ion components,
\begin{eqnarray}
  {\cal C}_{e\smI}^\smGT
  &=&
  \frac{\kappa_e^2}{2\pi}\, \omega_\smI^2\,
  \left(\frac{\beta_e m_e}{2\pi}\right)^{1/2}
  \frac{\Gamma(3/2)}{\Gamma(\nu/2)}\,
  \frac{1}{2}\,
  \left(\frac{4}{\lambda_e^2}\right)^{(\nu-3)/2}
  \Gamma\left( \frac{\nu-3}{2}\right) 
\label{dedteiRR}
\end{eqnarray}

\subsection{\label{sec:LBE} Lenard-Balescu Equation: 
Long-Distance Physics}

We now wish to calculate the leading order long-distance
physics by working in spatial dimensions $\nu<3$. The rate
of energy exchange from the electrons to the ions is
\begin{eqnarray}
  \frac{\partial{\cal E}_e^\smLT}{\partial t}
  =
  2\! \int\! \frac{d^\nu p_e}{(2\pi\hbar)^\nu}\, 
  \frac{p_e^2}{2 m_e}\,
  \frac{\partial f_e ({\bf p}_e)}{\partial t} \ ,
\label{dedteigtA}
\end{eqnarray}
and from (\ref{dedteigtAA}) we find the rate of energy exchange between
a spatially uniform distribution of electrons and ion species $i$, 
\begin{eqnarray}
\nonumber
  \frac{\partial{\cal E}_{e i}^\smLT}{\partial t}
  &=&
  -2\!\int\! \frac{d^\nu p_e}{(2\pi\hbar)^\nu}\, 
  \frac{d^\nu p_i}{(2\pi\hbar)^\nu}\,\frac{d^\nu k}{(2\pi)^\nu}
  \frac{p_e^2}{2 m_e}\,
  \, {\bm\nabla}_{\!p_e}\!\cdot\!{\bf k}\,
  \bigg\vert 
  \frac{e\, e_i}{k^2\, \epsilon({\bf k},{\bf k}\!\cdot\!{\bf v}_e)}
  \bigg\vert^2 
  \pi\delta(\hat{\bf k}\!\cdot\!{\bf v}_e-\hat{\bf k}\!\cdot\!{\bf v}_i)
\\[5pt] 
  && \hskip7cm 
  \Big(
  \hat{\bf k}\!\cdot\!{\bf v}_i\, \beta_i
  -
  \hat{\bf k}\!\cdot\!{\bf v}_e\,\beta_e
  \Big)
  f_e({\bf p}_e)f_i({\bf p}_i) \ ,
\label{dedteigtB}
\end{eqnarray}
where we have used the distribution (\ref{defFiA}) to write
${\bm\nabla}_{\!p_b} f_b({\bf p}_b) = - \beta_b\, {\bf v}_b\, f_b({\bf
p}_b)$.  Integrating ${\bf p}_e$ by parts using ${\bf k}\cdot {\mmbf
\nabla}_{ \!p_e}(p_e^2/2m_e)={\bf k}\cdot{\bf v}_e$, and integrating
over the ion distribution with (\ref{intdelfi}) gives 
\begin{eqnarray}
  \frac{\partial{\cal E}_{e i}^\smLT}{\partial t}
  &=&
  2\Big(\frac{\beta_e}{\beta_i}  - 1\Big)\!\!\! 
  \int\!\! \frac{d^\nu p_e}{(2\pi\hbar)^\nu}\, 
  \frac{d^\nu k}{(2\pi)^\nu}
  \frac{\pi\,e^2\,{\bf k}\!\cdot\!{\bf v}_e \,
  \rho_i(\hat{\bf k}\cdot{\bf v}_e) }{\big\vert
  k^2 + F(\hat{\bf k}\!\cdot\!{\bf v}_e)\big\vert^2}\, 
  f_e({\bf p}_e) \ ,
\label{dedteigtD}
\end{eqnarray}
where we have used (\ref{epsF}) and (\ref{rhocdef}).  When the
electron and ion temperatures are equal, $\beta_e=\beta_i$, we see
that the rate vanishes, as it must. Inserting unity in the form
\begin{eqnarray}
  \int_{-\infty}^\infty\! dv\, 
  \delta(v - \hat{\bf k}\!\cdot\!{\bf v}_e) =1 
\end{eqnarray}
allows us to express the rate as 
\begin{eqnarray}
  \frac{\partial{\cal E}_{e i}^\smLT}{\partial t}
  \!=\!
  2\Big(\frac{\beta_e}{\beta_i}-1\Big)\!\!
  \int_{-\infty}^\infty \!\! dv \!
  \int\!\! \frac{d^\nu p_e}{(2\pi\hbar)^\nu}\,
  \frac{d^\nu k}{(2\pi)^\nu}\,
  \frac{\pi e^2 k v \rho_i(v)}
  {\big\vert k^2 \!+\! F(v)\big\vert^2}\,
  \delta(v \!-\! \hat{\bf k}\!\cdot\!{\bf v}_e) 
  f_e({\bf p}_e) \ ,
\label{dedteigtG}
\end{eqnarray}
and upon performing the electron momentum integrals with
(\ref{intdelfi}), we find
\begin{eqnarray}
  \frac{\partial{\cal E}_{e i}^\smLT}{\partial t}
  \!=\!
  2\,\kappa_e^2
  \left(\frac{\beta_e m_e}{2\pi} \right)^{1/2} \!\!
  \Big(T_\smI - T_e\Big)\,
  \pi \int\!  \frac{d^\nu k}{(2\pi)^\nu}\,k
  \int_{-\infty}^\infty \!\! dv\,
  \frac{v \rho_i(v)}
  {\big\vert k^2 \!+\! F(v)\big\vert^2}\,
  e^{-\frac{1}{2}\,\beta_e m_e v^2} \ ,
\label{dedteigtH}
\end{eqnarray}
where we have used (\ref{kbdef}) for $\kappa_e$.

For individual ion species $i$, the $v$-integral can only
be performed numerically; however, if we sum over all ion
species we can perform the integral by completing a contour
in the complex $v$-plane~\cite{bps}: 
\begin{eqnarray}
  \lim_{m_e \to 0} {\sum}_i
  \int_{-\infty}^\infty \!\! dv\,
  \frac{v \rho_i(v)}
  {\big\vert k^2 \!+\! F(v)\big\vert^2}\,
  e^{-\frac{1}{2}\,\ell \beta_e m_e v^2}
  =  
  \frac{\omega_\smI^2}{(k^2 + \kappa_e^2)^2}
  ~~~~\text{with}~~ \omega_\smI^2 =  {\sum}_i\omega_i^2 \ .
\label{intrhoI}
\end{eqnarray}
Summing over the ions in (\ref{dedteigtH}) therefore allows us
to express 
\begin{eqnarray}
  \frac{\partial{\cal E}_{e \smI}^\smLT}{\partial t}
  \!=\! 
  2\,\kappa_e^2\,\omega_\smI^2
  \left(\frac{\beta_e m_e}{2\pi} \right)^{1/2} \!\!
  \Big(T_\smI - T_e\Big)\,
  \pi \int\!  \frac{d^\nu k}{(2\pi)^\nu}\,\frac{k}{(k^2 +
  \kappa_e^2)^2} \ .
\label{dedteigtJ}
\end{eqnarray}
Note that the $k$-integral 
\begin{eqnarray}
  \mathbb{J} 
  =
  \int \!\! \frac{d^\nu k}{(2\pi)^\nu}
  \frac{k}{(k^2 + \kappa_e^2)^2} 
\label{intke}
\end{eqnarray}
converges both in the UV and IR. This is because the $\kappa_e^2$ term
in the denominator provides a long-distance cutoff (small values of
$k$), while the condition $\nu<3$ provides short-distance convergence
(large values of $k$). This integral can be performed by first
converting to hyperspherical coordinates and then changing variables
by $k^2 = \kappa_e^2\,t$:
\begin{eqnarray}
  \mathbb{J} 
  =
  \frac{\Omega_{\nu-1}}{(2\pi)^\nu}
  \int_0^\infty \! k^{\nu-1} dk\,
  \frac{k}{(k^2 + \kappa_e^2)^2} 
  =
  \frac{\kappa_e^{\nu-3}}{2}\,  \frac{\Omega_{\nu-1}}{(2\pi)^\nu}
  \int_0^\infty \! dt\, t^{\nu-1/2} (t+1)^{-2} \ ,
\label{intkeB}
\end{eqnarray}
where the solid angle integrals $\Omega_{\nu-1}$ are given in
(\ref{eintGamma}). The $t$-integral can be recognized as the Beta 
function, defined in (\ref{Bxyanother})
with $x=(\nu+1)/2$ and $y=(3-\nu)/2$. Inserting the appropriate
Gamma functions and factors of $\pi$ gives 
\begin{eqnarray}
   \int\!\!\frac{d^\nu k}{(2\pi)^\nu}\, \frac{k}{(k^2 + \kappa_e^2)^2} 
  &=&
  \frac{1}{4\pi^2}\, 
  \frac{\Gamma(3/2)}{\Gamma(\nu/2)}\, 
  \Gamma\!\left(\frac{\nu+1}{2}\right)
  \Gamma\!\left(\frac{3-\nu}{2}\right)
  \left(\frac{\kappa_e^2}{4\pi}\right)^{(\nu-3)/2} \ ,
\label{JevalSimp}
\end{eqnarray}
where $\Gamma(3/2)=\sqrt{\pi}/2$.  Substituting (\ref{JevalSimp}) back
(\ref{dedteigtJ}) gives 
\begin{eqnarray}
  \frac{\partial{\cal E}_{e \smI}^\smLT}{\partial t}
  \!=\!
  2\,\kappa_e^2\,\omega_\smI^2
  \left(\frac{\beta_e m_e}{2\pi} \right)^{1/2} \!\!
  \Big(T_\smI - T_e\Big)\,
  \frac{1}{4\pi}\, 
  \frac{\Gamma(3/2)}{\Gamma(\nu/2)}\, 
  \Gamma\!\left(\frac{\nu+1}{2}\right)
  \Gamma\!\left(\frac{3-\nu}{2}\right)
  \left(\frac{\kappa_e^2}{4\pi}\right)^{(\nu-3)/2} ,
\label{dedteigtK}
\end{eqnarray}
or the rate coefficient 
\begin{eqnarray}
  {\cal C}_{e \smI}^\smLT
  \!=\!
  \frac{\kappa_e^2}{2\pi}\,\omega_\smI^2
  \left(\frac{\beta_e m_e}{2\pi} \right)^{1/2} 
  \frac{\Gamma(3/2)}{\Gamma(\nu/2)}\, 
  \frac{1}{2}\,
  \Gamma\!\left(\frac{\nu+1}{2}\right)
  \Gamma\!\left(\frac{3-\nu}{2}\right)
  \left(\frac{\kappa_e^2}{4\pi}\right)^{(\nu-3)/2} .
\label{dedteigtL}
\end{eqnarray}
\subsection{Combining the Leading and Next-to-Leading Order Terms}

Recall that the rate coefficient in $\nu<3$ and $\nu>3$
takes the exact analytic form 
\begin{eqnarray}
  {\cal C}_{e \smI}^\smGT
  &=&
  \frac{\kappa_e^2}{2\pi} ~\omega_\smI^2 ~
  \left(\frac{\beta_e m_e}{2\pi}\right)^{1/2}
  \frac{\Gamma(3/2)}{\Gamma(\nu/2)}\,
  \frac{1}{2}\,\Gamma\left( \frac{\nu-3}{2}\right)\,
  \left(\frac{4}{\lambda_e^2}\right)^{(\nu-3)/2}
\label{CeIlt}
\\[8pt]
  {\cal C}_{e \smI}^\smLT
  &=&
  \frac{\kappa_e^2}{2\pi}\,
  \omega_\smI^2\, 
  \left(\frac{\beta_e m_e}{2\pi}\right)^{1/2}
  \frac{\Gamma(3/2)}{\Gamma(\nu/2)}\, 
  \frac{1}{2}\,
  \Gamma\!\left(\frac{\nu+1}{2}\right)
  \Gamma\!\left(\frac{3-\nu}{2}\right)\,
  \left(\frac{\kappa_e^2}{4\pi}\right)^{(\nu-3)/2} \ .
\label{CeIgt}
\end{eqnarray}
The gamma functions in the above expressions can be expanded in the
parameter $\epsilon=\nu-3$:
\begin{eqnarray}
  \Gamma(\epsilon) 
  &=&
  \frac{1}{\epsilon} - \gamma + {\cal O}(\epsilon) 
\label{gammaA}
\\[5pt]
  \Gamma(1+\epsilon) 
  &=&
  1 - \gamma\,\epsilon + {\cal O}(\epsilon^2) \ .
\label{gammaB}
\end{eqnarray}
We often need to multiply a term of the form $A^{\nu-3}$ by a pole
$1/(\nu-3)$. This will produce a pole term {\em and} a finite
contribution. In fact, this is the origin of the coefficient under the
logarithm, and since this point is so important, I will reiterate it
once again. For ease of notation let $\epsilon=\nu-3$, so that the
$\nu\to 3$ limit is the same as the $\epsilon\to 0$ limit. In any
calculation we can therefore drop terms ${\cal O}(\epsilon)$ and
higher; however, we must be careful and drop such terms too soon. This
is because an order $\epsilon$ term could multiply a pole term of the
form $1/\epsilon$, thereby giving a finite nonzero result in the limit
$\epsilon \to 0$. The following example illustrates this point. Let
us consider the product of $A^\epsilon$ with the pole $1/\epsilon$.
We first expand $A^\epsilon$ in powers of $\epsilon$ as follows
\begin{eqnarray}
  A^\epsilon 
  = 
  \exp\{ \ln A^\epsilon \} 
  = 
  \exp\{ \epsilon\,\ln A \}
  =
  1 + \epsilon \, \ln A + {\cal O}(\epsilon^2) \ .
\end{eqnarray}
Upon multiplying this expression by the pole we find
\begin{eqnarray}
  \frac{A^\epsilon}{\epsilon}
  =
  \frac{1}{\epsilon} + \ln A + {\cal O}(\epsilon) \ .
\label{polemult}
\end{eqnarray}
Therefore, upon using expression (\ref{polemult}) in (\ref{CeIlt})
and (\ref{CeIgt}), we find
\begin{eqnarray}
  \Gamma\left(\frac{\nu-3}{2}\right)
  \left[\frac{4}{\lambda_e^2}\right]^{(\nu-3)/2}
  &\!\!=\!&
  \phantom{-}
  \frac{2}{\nu-3} 
  + 
  \ln\!\left\{ \frac{4}{\lambda_e^2}\right\}   
  -
  \gamma  
\\[5pt]
  \Gamma\left(\frac{\nu+1}{2}\right)
  \Gamma\left(\frac{3-\nu}{2}\right) 
  \left[\frac{\kappa_e^2}{4\pi}\right]^{(\nu-3)/2} 
  &\!\!=\!&
  -\frac{2}{\nu-3} 
  - 
  \ln\!\left\{\frac{\kappa_e^2}{4\pi}\right\} 
  -
  1 \ .
\end{eqnarray}
This gives the rate coefficient
\begin{eqnarray}
  {\cal C}_{e \smI}
  =
  {\cal C}_{e \smI}^\smGT 
  +
  {\cal C}_{e \smI}^\smLT
  =
  \Bigg[
  \frac{\kappa_e^2}{2\pi} ~\omega_\smI^2 ~
  \left(\frac{\beta_e m_e}{2\pi}\right)^{1/2}
  \Bigg]
  \cdot
  \frac{1}{2}
  \Bigg[
  \ln\left\{\frac{16 \pi}{\kappa_e^2\,\lambda_e^2} \right\} -\gamma -1
  \Bigg] \ .
\end{eqnarray}
Note that the pole terms have canceled, rendering a finite result
accurate to leading and next-to-leading order in $g$.  As we have
seen, the next-to-leading order term gives the exact coefficient under
the logarithm (including the term $-\gamma -1$). The argument of the
logarithm can be expressed as
\begin{eqnarray}
  \frac{16\pi}{\lambda_e^2\, \kappa_e^2}
  =
  \frac{8\,m_e T_e}{\hbar^2}\,\frac{T_e}{e^2\, n_e}
  =
  \frac{8\, T_e^2}{\hbar^2\, \omega_e^2} \ ,
\end{eqnarray}
and therefore the rate coefficient takes the form
\begin{eqnarray}
  {\cal C}_{e\smI}
  &=& 
  \frac{\kappa_e^2}{2\pi}\, \omega_\smI^2\,
  \sqrt{\frac{\beta_e m_e}{2\pi}}\, \ln\Lambda_\smBPS 
\label{bpsrateA}
\\[10pt]
  \ln\Lambda_\smBPS
  &=&
  \frac{1}{2}\left[\ln\!\left\{\frac{8 T_e^2}{\hbar^2 \omega_e^2}
  \right\} - \gamma - 1 \right] \ .
\label{bpsrateB}
\end{eqnarray}

\end{document}